\documentclass[useAMS,usenatbib]{mnras}

\usepackage[english]{babel}
\usepackage{amsmath,amssymb}
\usepackage{graphicx, subfig}
\usepackage[normalem]{ulem}

\usepackage{verbatim}

\usepackage{color}
\usepackage{multirow}
\usepackage{mathtools}
\usepackage{epstopdf}

\usepackage{url}
\usepackage{color}

\def\nat{Nature}

\def\mnras{MNRAS}
\def\apj{ApJ}
\def\apjl{ApJL}
\def\apjs{ApJS}
\def\aap{A\&A}
\def\aaps{A\&A Supp.}

\def\araa{ARA\&A}

\def\pasp{Publ. Astr. Soc. Pac.}

\def\nar{New Astronomy Reviews}

\def\be{\begin{equation}} 
\def\ee{\end{equation}} 
\def\ba{\begin{eqnarray}} 
\def\ea{\end{eqnarray}}

\def\cc{\,{\rm {cm^{-3}}}} 
\def\msun{{\Msun}}

\def\HH{${\rm {H_2}}$}

\def\HI{\hbox{H~$\scriptstyle\rm I\ $}}

\def\gsim{\lower.5ex\hbox{\gtsima}} 
\def\lsim{\lower.5ex\hbox{\ltsima}} \def\gtsima{$\; \buildrel > \over 
\sim \;$} \def\ltsima{$\; \buildrel < \over \sim \;$} \def\prosima{$\; 
\buildrel \propto \over \sim \;$} \def\gsim{\lower.5ex\hbox{\gtsima}} 
\def\lsim{\lower.5ex\hbox{\ltsima}} 
\def\simgt{\lower.5ex\hbox{\gtsima}} 
\def\simlt{\lower.5ex\hbox{\ltsima}} 
\def\simpr{\lower.5ex\hbox{\prosima}} \def\la{\lsim} \def\ga{\gsim} 
  
 \def\gtsima{$\; \buildrel > \over \sim \;$} 
\def\ltsima{$\; \buildrel < \over \sim \;$} 
\def\gsim{\lower.5ex\hbox{\gtsima}} 
\def\lsim{\lower.5ex\hbox{\ltsima}} 
\def\simgt{\lower.5ex\hbox{\gtsima}} 
\def\simlt{\lower.5ex\hbox{\ltsima}} 
\def\simpr{\lower.5ex\hbox{\prosima}} 
\def\la{\lsim} 
\def\ga{\gsim}

\def\msun{\,{\rm \Msun}}

\def\E3{{\cal E}_{\rm g}^{III}}

\def\r12{r_{1/2}} 
\def\x12{x_{1/2}} 
\def\v12{v_{1/2}}

\def\msun{{\rm M}_{\odot}}
\def\zsun{{\rm Z}_{\odot}}

\def\HI{\hbox{H~$\scriptstyle\rm I $~}}

\def\CII{\hbox{[C~$\scriptstyle\rm II $]}}
\def\CIIion{\hbox{C~$\scriptstyle\rm II $}}

\newcommand\textlcsc[1]{\textsc{\MakeLowercase{#1}}}
\newcommand{\quotes}[1]{``#1''}

\def\angstrom{\textrm{A\kern -1.3ex\raisebox{0.6ex}{$^\circ$}}}
\def\myr{\rm Myr}

\defcitealias{krumholz:2009apj}{KTM09}
\defcitealias{pallottini:2014_sim}{P14}
\defcitealias{ostriker:1988rvmp}{OM88}
\defcitealias{weaver:1977apj}{W77}
\defcitealias{semenov:2015}{S16}

\def\lsun{{\rm L}_{\odot}}
\def\cc{{\rm cm}^{-3}}
\def\surfd{\msun\,{\rm kpc}^{-2}}
\def\surfl{\lsun\,{\rm kpc}^{-2}}

\begin{document}

\date{}
\pagerange{\pageref{firstpage}--\pageref{lastpage}} \pubyear{2016}
\title[The internal structure of $z\simeq6$ galaxies]{Zooming on the internal structure of $z\simeq6$ galaxies}
\author[Pallottini et al.]{A. Pallottini$^{1,2,3}$\thanks{\href{mailto:andrea.pallottini@sns.it}{andrea.pallottini@sns.it}; \href{mailto:ap926@cam.ac.uk}{ap926@cam.ac.uk}},
A. Ferrara$^{3,4}$, S. Gallerani$^{3}$, L. Vallini$^{5,6}$, R. Maiolino$^{1,2}$, S. Salvadori$^{7}$\\
$^{1}$Cavendish Laboratory, University of Cambridge, 19 J. J. Thomson Ave., Cambridge CB3 0HE, UK\\
$^{2}$Kavli Institute for Cosmology, University of Cambridge, Madingley Road, Cambridge CB3 0HA, UK\\
$^{3}$Scuola Normale Superiore, Piazza dei Cavalieri 7, I-56126 Pisa, Italy\\
$^{4}$Kavli IPMU, The University of Tokyo, 5-1-5 Kashiwanoha, Kashiwa 277-8583, Japan\\
$^{5}$Dipartimento di Fisica e Astronomia, Universit\'{a} di Bologna, viale Berti Pichat 6/2, I-40127 Bologna, Italy\\
$^{6}$INAF, Osservatorio Astronomico di Bologna, via Ranzani 1, I-40127 Bologna, Italy\\
$^{7}$GEPI, Observatoire de Paris, Place Jules Jannsen, 92195 Meudon, Paris, France
}

\maketitle

\label{firstpage}

\begin{abstract}
We present zoom-in, AMR, high-resolution ($\simeq 30\, {\rm pc}$) simulations of high-redshift ($z \simeq 6$) galaxies with the aim of characterizing their internal properties and interstellar medium. Among other features, we adopt a star formation model based on a physically-sound molecular hydrogen prescription, and introduce a novel scheme for supernova feedback, stellar winds and dust-mediated radiation pressure. In the zoom-in simulation the target halo hosts \quotes{Dahlia}, a galaxy with a stellar mass $M_{\star}=1.6\times 10^{10}\msun$, representative of a typical $z\sim 6$ Lyman Break Galaxy.
{Dahlia} has a total \HH~mass of $10^{8.5}\msun$, that is mainly concentrated in a disk-like structure of effective radius $\simeq 0.6$ kpc and scale height $\simeq 200$ pc. Frequent mergers drive fresh gas towards the centre of the disk, sustaining a star formation rate per unit area of $\simeq 15\,\msun\,{\rm yr}^{-1}\,{\rm kpc}^{-2}$. The disk is composed by dense ($n \gsim 25\,\cc$), metal-rich ($Z \simeq 0.5\,\zsun$) gas, that is pressure-supported by radiation.
We compute the $158\mu$m \CII~emission arising from {Dahlia}, and find that $\simeq 95\%$ of the total \CII~luminosity ($L_{\rm [CII]}\simeq10^{7.5}\lsun$) arises from the \HH~disk. Although $30\%$ of the \CIIion~mass is transported out of the disk by outflows, such gas negligibly contributes to \CII~emission, due to its low density ($n \lsim 10\,\cc$) and metallicity ($Z\lsim 10^{-1}\zsun$). {Dahlia} is under-luminous with respect to the local \CII-$SFR$ relation; however, its luminosity is consistent with upper limits derived for most $z\sim6$ galaxies.
\end{abstract}


\section{Introduction}

The discovery and characterization of primeval galaxies constitute some of the biggest challenges in current observational and theoretical cosmology\footnote{In the following we assume cosmological parameters compatible with \emph{Planck} results, i.e. a $\Lambda$CDM model with total matter, vacuum and baryonic densities in units of the critical density $\Omega_{\Lambda}= 0.692$, $\Omega_{m}= 0.308$, $\Omega_{b}= 0.0481$, Hubble constant $\rm H_0=100\,{\rm h}\,{\rm km}\,{\rm s}^{-1}\,{\rm Mpc}^{-1}$ with ${\rm h}=0.678$, spectral index $n=0.967$, $\sigma_{8}=0.826$ \citep[][]{planck:2013_xvi_parameters}.}.

Deep optical/near infrared (IR) surveys \citep{Dunlop13,Madau14,Bouwens:2015} have made impressive progresses in identifying galaxies well within the Epoch of Reionization ($z\simeq6$). Such surveys yield key information about the star formation (SF) of hundreds of galaxies in the early Universe. They also allow to statistically characterize galaxies in terms of their UltraViolet (UV) luminosity up to $z\sim10$ \citep{Bouwens:2015}. However -- using these surveys broad band alone -- little can be learned about other properties as their gas and dust content, metallicity, interactions with the surrounding environment \citep[e.g.][]{Barnes:2014PASP}, feedback \citep[e.g.][]{Dayal14}, and outflows \citep{gallerani:2016outflow}. 

To obtain a full picture of these systems, optical/IR surveys must be complemented with additional probes. Information on the metal content and energetics of the interstellar medium (ISM) can be obtained with observations of Far IR (FIR) fine structure lines, and in particular the \CII~{\small$\left(^{2}P_{3/2} \rightarrow\,^{2}P_{1/2}\right)$} line at 157.74~$\mu$m. The \CII~line is the dominant coolant of the ISM being excited in different ISM phases, as the diffuse cold neutral medium (CNM), warm neutral medium (WNM), high density photodissociation regions (PDRs), and -- to a lower extent -- ionized gas \citep[][]{Tielens:1985ApJ,Wolfire:1995ApJ,Abel:2006MNRAS,Vallini:2013MNRAS}. As \CII~emission can be enhanced by shocks, it has been suggested as a good outflow tracer\\ (e.g. \citealt{maiolino:2012,kreckel:2014apj,cicone:2015aa,janssen:2016arxiv}), and can thus in general be used to study feedback processes in galaxies.

Observationally, the \CII~line is a promising probe as it is often the brightest among FIR emission lines, accounting for up to $\sim1\%$ of the total IR luminosity of galaxies \citep[e.g.][]{Crawford:1985ApJ,Madden:1997ApJ}. It has been successfully used to probe the low-$z$ ISM \citep[e.g.][]{delooze:2014aa}. The unprecedented sensitivity of the Atacama Large Millimeter/Submillmeter Array (ALMA) makes it possible for the first time to use \CII~emission to characterize high-$z$ galaxies. Before the ALMA advent, in fact, detections were limited to a handful of QSO host galaxies, and rare galaxies with extreme SF rates \citep[$SFR\simeq10^3\msun\,{\rm yr}^{-1}$, e.g.][]{maiolino:2005AA,debreuck:2011,Carilli:2013ARA&A,gallerani:2012aa,cicone:2015aa}.

%
%
However, for \quotes{normal} star forming galaxies ($\lsim10^{2}\msun\,{\rm yr}^{-1}$) at $z\sim 6-7$ early ALMA searches for \CII~lines have mostly yielded upper limits (e.g. \citealt{ouchi2013} \citealt{kanekar2013}; \citealt{ota:2014apj,schaerer:2015}). The situation has changed recently with a number of robust \CII~detections (e.g. \citealt{maiolino:2015arxiv,capak:2015arxiv}; \citealt{Willott:2015arXiv15,knudsen:2016arxiv}).

In many cases the high-$z$ \CII~line luminosity is fainter than expected from the \CII-$SFR$ relation found in local galaxies \citep{delooze:2014aa}. To explain such \CII-$SFR$~\emph{deficit}, some efforts have been devoted to model the \CII~emission from high-$z$ galaxies \citep{nagamine:2006ApJ,Vallini:2013MNRAS,munoz:2014MNRAS,vallini:2015,olsen:2015apj}. In brief, these theoretical works show that the \CII-$SFR$~deficit can be ascribed to different effects:
\begin{itemize}
\item[(a)] Lower metallicity of high-$z$ galaxies \citep{Vallini:2013MNRAS,munoz:2014MNRAS,vallini:2015}, in particular supported by observations of lensed galaxies \citep{knudsen:2016arxiv}.
\item[(b)] Suppression of \CII~line around star forming regions \citet{Vallini:2013MNRAS}, typically observed as a displacement of the \CII~ with respect to the UV emitting region, as seen e.g. in BDF3299 \citep{maiolino:2015arxiv} and in some of the \citet{capak:2015arxiv} galaxies. This would be a signature of stellar feedback heating/ionizing the putative \CII-emitting gas.
\item[(c)] Suppression of \CII~line by the increased CMB temperature in the WNM/CNM component \citep[][]{pallottini:2015_cmb,vallini:2015}, similarly to what observed for dust emission \citep{dacunha:2013apj}.
\end{itemize}

Simulating the ISM of early galaxies at sufficient resolution and including feedback effects might shed light on these questions. Feedback prescriptions are particularly important as such process regulates the amount of (dense) gas likely to radiate most of the power detected with FIR lines. Several studies have explored optimal strategies to include feedback in galaxy simulations.

For some works, the interest is in the comparison between different kind of stellar feedback prescription, as modelled via thermal and/or kinetic energy deposition in the gas from supernovae (SN), winds \citep[][]{agertz:2012arxiv,fire:2014mnras,barai:2015mnras,agertz:2015apj}, and radiation pressure \citep[][]{wise:2012radpres,ceverino:2014}; other analyses focus on implementing complex chemical networks in simulations \citep{tomassetti:2015MNRAS,maio:2015,bovino:2015arxiv,richings:2016,grassi_dust:2016}, radiative transfer effect \citep{petkova:2012mnras,roskar:2014,rosdahl:2015mnras,maio:2016mnras}, or aim at removing tensions between different coding approaches \citep[][]{agora:2013arxiv}.

%
Thus, we can improve galaxy simulations by providing theoretical expectations for \CII~that should be compared with state-of-the-art data. Such a synergy between theory and observations, in turn, can guide the interpretation of upcoming ALMA data and drive future experiments of large
scale \CII~mapping \citep{Gong:2012ApJ,silva:2015apj,bin:2015mapping, pallottini:2015_cmb}, which would led to a statistical characterization of the high-$z$ galaxy population. In the present work we simulate a $z\sim6$ galaxy typically detected in \CII~with ALMA current observations.

The paper is structured as follows. In Sec. \ref{sec_numerical} we detail the numerical model used to set-up the zoom-in simulation, and describe the adopted \HH~star formation prescription (Sec. \ref{sec_model_sf}), mass and energy inputs from the stellar populations (Sec. \ref{sec_stellar_inputs}) and feedback (including SN, winds and radiation pressure Sec. \ref{sezione_blast} -- see also App. \ref{app_rad_press} and App. \ref{app_blastwave}). The results are discussed in Sec. \ref{sec_result}, where we analyze star formation history and feedback effects in relation to ISM thermodynamics (Sec. \ref{sec_sfr_result}) and its structural properties. The expected \CII~emission and other observational properties of high-$z$ galaxies are discussed in Sec. \ref{sec_final_results}. Conclusions are given in Sec. \ref{sec_conclusioni}.

%

\section{Numerical simulations}\label{sec_numerical}

\begin{table}
\centering
\begin{tabular}{ccccccc}
\hline
~ & $m_{dm}$ & $m_{b}$ & $\Delta_{x}^{\rm max}$ & $\Delta_{x}^{\rm min}$ & $\Delta_{x}^{\rm min}$ at $z=6$\\
~ & \multicolumn{2}{c}{$\msun/{\rm h}$} & \multicolumn{2}{c}{${\rm kpc}/{\rm h}$} & pc\\
\hline
{\tt cosmo} & $3.4\times 10^{7}$ & $-$                & $78.1$ & $78.1$ & $2.5\times10^3$\\
{\tt zoom}  & $6.7\times 10^{4}$ & $1.2\times 10^{4}$ & $9.7$  & $0.1$  & $32.1$\\
\end{tabular}
\caption{Resolution set-up for the cosmological run ({\tt cosmo}) and subsequent zoom-in ({\tt zoom}) simulation. $m_{dm}$ and $m_{b}$ are in units of $\msun/{\rm h}$ and indicate the dark matter (DM) and baryon mass resolution, respectively; $\Delta_{x}^{\rm max}$ and $\Delta_{x}^{\rm min}$ indicate the coarse grid and minimum available refinement scale, respectively. Both scales are reported in comoving ${\rm kpc}/{\rm h}$. For $\Delta_{x}^{\rm min}$ we also report also the physical pc scale at $z=6$. For the {\tt cosmo} run, no refinement is used, and for the {\tt zoom}, we indicate the increased resolution of the zoomed halo due to the multi-mass approach and the AMR.
\label{tagella_res}}
\end{table}

We carry out our simulation using a customized version of the adaptive mesh refinement (AMR) code \textlcsc{ramses} \citep[][]{Teyssier:2002}. \textlcsc{ramses} is an octree-based code that uses Particle Mesh N-body solver for the dark matter (DM) and an unsplit 2nd-order MUSCL\footnote{MUSCL: Monotone Upstream-centred Scheme for Conservation Laws} scheme for the baryons. Gravity is accounted by solving the Poisson equation on the AMR grid via a multi-grid scheme with Dirichlet boundary conditions on arbitrary domains \citep{guillet:2011Jcoph}. For the present simulation we choose a refinement based on a Lagrangian mass threshold-based criterion.

Chemistry and heating/cooling processes of the baryons are implemented with \textlcsc{grackle} 2.1\footnote{See also \url{https://grackle.readthedocs.org/}} \citep{bryan:2014apjs}, the standard library of thermo-chemical processes of the {\tt AGORA} project \citep{agora:2013arxiv}. Via \textlcsc{grackle}, we follow the \hbox{H}~and \hbox{He}~primordial network and tabulated metal cooling and photo-heating rates calculated with \textlcsc{cloudy} \citep{cloudy:2013}. Cooling includes also inverse Compton off the cosmic microwave background (CMB), and heating from a redshift-dependent ionizing UV background \citep[][UVB]{Haardt:2012}. Since \HH~gas phase formation is not accounted for, we do not include the cooling contribution of such species.

Because of stellar feedback (Sec \ref{sec_stellar_inputs} and \ref{sezione_blast}), the gas can acquire energy both in thermal and kinetic form. The distinction is considered by following the gas evolution of the standard thermal energy and a \quotes{non-thermal} energy \citep{agertz:2012arxiv}. Such approach is one of the possible scheme used to solve the over-cooling problem that affect galaxy-scale simulations \citep[see][ and references therein]{dale:2015new}. The non-thermal energy mimics turbulence, i.e. it is not affected by cooling. The non-thermal energy variation is due to gas advection ($v\nabla v$), work ($PdV$), and dissipation \citep{agertz:2015apj}. Following \citet{maclow1999turb} we assume a dissipation time scale proportional to the size of the cell (injection scale) and inversely proportional to the Mach number\footnote{While the distinction in thermal and non-thermal is similar to previous works \citep[e.g.][]{agertz:2015apj}, we note that usually the time scale for dissipation is fixed to $10\,\myr$.}. Since the dynamical time is essentially set by the free-fall time, the dissipation time can be written as $t_{\rm diss} = 9.785 (l_{\rm cell}/100\,{\rm pc})/(v_{\rm turb}/10\,{\rm km}\,{\rm s}^{-1}) \myr$. Then, the non-thermal energy loss due to dissipation can be written as $\dot{e}_{\rm nth} = -e_{\rm nth}/t_{\rm diss}$ \citep[][see eq. 2]{teyssier:2013mnras}. As noted in \citet{teyssier:2013mnras}, such scheme for non-thermal energy and its dissipation gives results qualitatively similar to a delayed cooling approach \citep{stinson:2006mnras}.

\subsection{Initial conditions}

The initial conditions (IC) used for the suite are generated with \textlcsc{music} \citep{hahn:2011mnras}. \textlcsc{music} produces IC on nested grid using a real-space convolution approach \citep[cf.][]{bertschinger:1995astro}. The adopted Lagrangian perturbation theory scheme is perfectly suited to produce IC for multi-mass simulations and -- in particular -- zoom simulations. To generate the ICs, the transfer functions are taken from \citep{eisenstein:1998apj}.

To set-up the zoom-in simulation, we start by carrying out a cosmological DM-only run. The simulation evolves a volume $V^{\rm cosmo}=(20\,{\rm Mpc}/{\rm h})^{3}$ from $z=100$ to $z=6$ with DM mass resolution of $m_{dm}^{\rm cosmo} = 3.4\times 10^{7} /{\rm h}\,\msun$. The resolution of the coarse grid is $\Delta x^{\rm cosmo} = 78.1 /{\rm h}\,{\rm kpc}$, and we do not include additional levels of refinement. Using \textlcsc{hop} \citep{eisenstein_hop_1998apj} we find the DM halo catalogue at $z=6$. The cumulative halo mass function extracted from the catalogue is in agreement with analytical expectations \citep[e.g.][]{sheth:1999mnras}, within the precision of halo-finder codes \citep[e.g.][]{knebe:2013arxiv}.

From the catalogue we select a halo with DM mass $M_{\rm h} \simeq 10^{11}/{\rm h}\,\msun$ (resolved by $\simeq5\times10^{4}$ DM particles), whose virial radius is $r_{\rm vir}\simeq 15\,{\rm kpc}$ at $z=6$. Using \textlcsc{hop} we select the minimum ellipsoid enveloping $10\,r_{\rm vir}$, and trace it back to $z=100$. As noted in \citet[][]{onorbe:2014mnras}, this is usually sufficient to avoid contamination\footnote{A posteriori, we have checked that the halos in the zoom-in region have a contamination level $\lsim0.1\%$.}. At $z=100$ the trace back volume is $V^{\rm zoom}\simeq(2.1\,{\rm Mpc}/{\rm h})^{3}$. Using \textlcsc{music} we recalculate the ICs, by generating 3 additional level of refinement. For such multi-mass set-up, the finer DM resolution is $m_{dm}^{\rm zoom} = 6.7\times 10^{4} /{\rm h}\,\msun$, that corresponds to a spatial resolution of $\Delta x^{\rm zoom} = 9.7 /{\rm h}\,{\rm kpc}$. We note that because of the traced back volume, our simulation is expected to probe not only the target halo, but also its satellites and environment, similar to other works (e.g. \citealt{fiacconi:2015}, where the target halo is chosen at $z\simeq 3$).

In the zoom-in simulation $\Delta x^{\rm zoom}$ corresponds to our coarse grid resolution, and we allow for 6 additional refinement levels, based on a Lagrangian mass threshold-based criterion. At $z=6$, the baryonic component of the selected halo has a mass resolution of $m_{b} = 1.8\times 10^{4}\msun$ and a physical resolution of $\Delta x^{\rm min} = 31.9\,{\rm pc}$. For convenience, a summary of the resolution outline can be found in Tab. \ref{tagella_res}. Note that the refined cell of our simulations have mass and size typical of molecular clouds \citep[MC, e.g.][]{gorti:2002apj,federrath:2013}.

In the present paper we refer to metallicity ($Z$) as the sum of all the heavy element species without differentiating among them, and assume solar abundance ratios \citep{asplund:2009ara&a}. In the IC, the gas is characterized by a mean molecular weight $\mu = 0.59$, and has metallicity floor $Z=Z_{\rm floor}>0$. The metallicity floor mimics the pre-enrichment of the halo at high-$z$, when we do not have the resolution to follow precisely star formation and gas enrichment. We set $Z_{\rm floor}=10^{-3}\zsun$, a level that is compatible with the metallicity found at high-$z$ in cosmological simulations for diffuse enriched gas \citep{dave:2011mnras,pallottini:2014_sim,maio:2015}. Note that such low metallicity only marginally affects the gas cooling time, but is above the critical metallicity for formation of Population III stars. Additionally, a posteriori, we have found that the metallicity floor contribute for only $\lsim 0.2\%$ of the total metal mass produced by stars by $z=6$ in the refined region.

\subsection{Star formation model}\label{sec_model_sf}

We model star formation (SF) by assuming a \HH~dependent Schmidt-Kennicutt relation \citep{schmidt:1959apj,kennicutt:1998apj}
\begin{figure}
\centering
\includegraphics[width=0.49\textwidth]{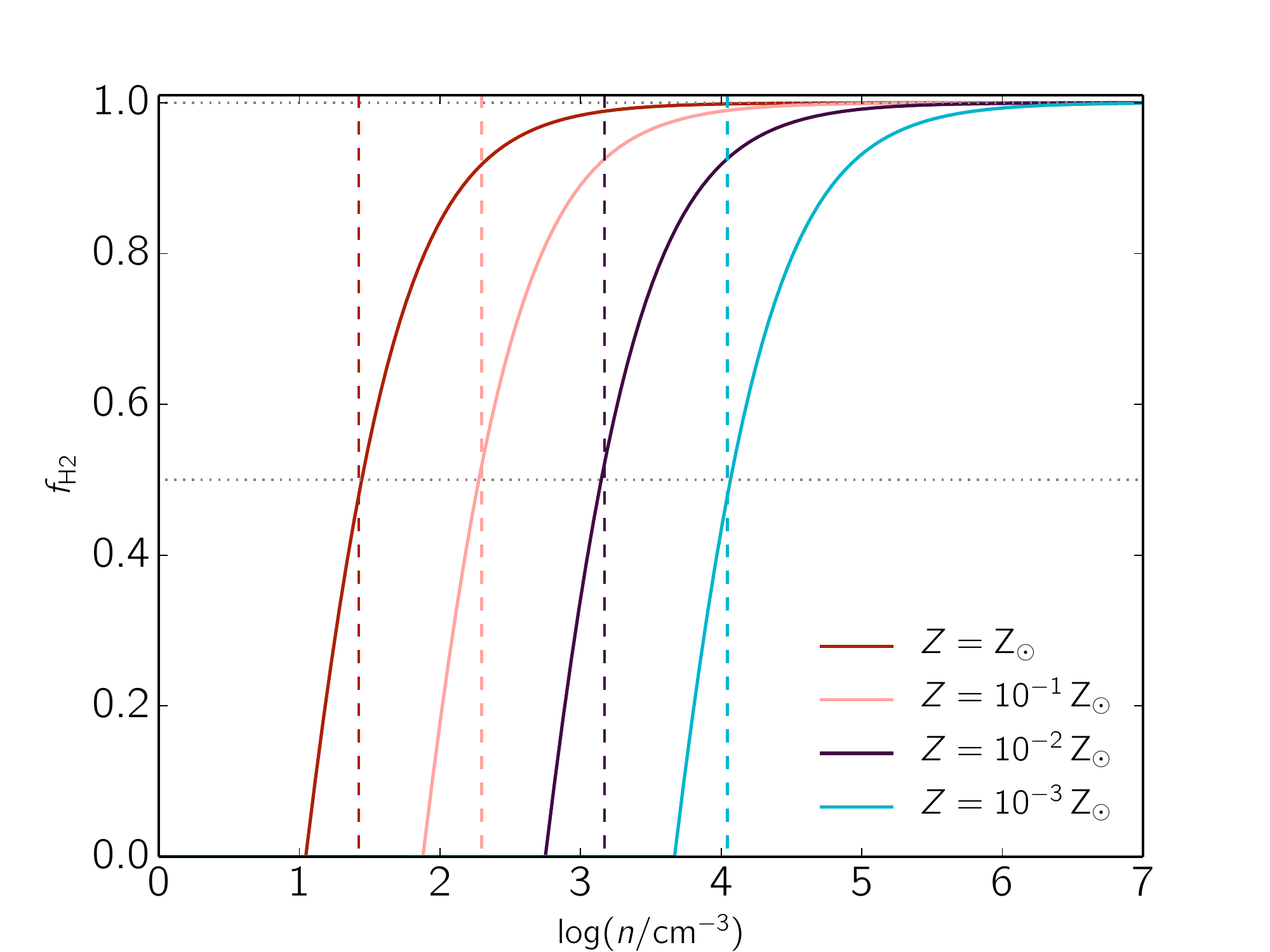}
\caption{
\HH~fraction ($f_{\rm H2}$) as a function of gas density ($n$) obtained using the \citetalias{krumholz:2009apj} model (eqs. \ref{eq_fh2_full}). Different solid lines correspond to different metallicity ($Z$) of the gas. Horizontal dotted grey lines mark $f_{\rm H2}$ values of $0.5$ and $1$. Vertical dashed lines indicate the critical density $n_{c}$ where $f_{\rm H2}=0.5$ for different $Z$; these critical density values are obtained as a fit (eq. \ref{eq_critical_density}) to the \citetalias{krumholz:2009apj} model (eq. \ref{eq_fh2_full}). See the text for details.
\label{fig_kmt_test}}
\end{figure}
\begin{subequations}\label{eq_sfr_tot}
\be\label{eq_sfr1}
\dot{\rho}_{\star}=f_{\rm H2} \rho/t_{\rm sf}\,
\ee
where $\dot{\rho}_{\star}$ is the local SF rate ($SFR$) density, $f_{\rm H2}$ the molecular hydrogen fraction, $\rho$ the gas density and $t_{\rm sf}$ the SF time scale. In eq. \ref{eq_sfr1} we assume the SF time scale to be proportional to the free-fall time, i.e.

\be\label{eq_sfr2}
t_{\rm sf} = \zeta_{\rm sf}^{-1} \sqrt{3\pi/(32\,G\rho)} \,,
\ee
\end{subequations}
where $\zeta_{\rm sf}$ describes the SF efficiency and it is treated as a parameter in the present work \citep[cf.][see discussion in Sec. \ref{sez_sfr_efficiency}]{semenov:2015}. To calculate $f_{\rm H2}$ we adopt the \citetalias{krumholz:2009apj} model \citep{krumholz:2008apj,krumholz:2009apj,mckee:2010apj}. Such model considers \HH~formation on dust grains by computing radiative transfer on a idealized MC and assumes equilibrium between formation and dissociation rate of \HH. The solution for $f_{\rm H2}$ can be approximated as

\begin{subequations}\label{eq_fh2_full}
\begin{align}
f_{\rm H2} &= \left(1 -0.75\,s/(1+0.25\,s) \right)\Theta(2-s)\\
s          &= \ln\left(1+0.6\,\chi +0.01\chi^{2}\right) /0.6\,\tau_{\rm uv}\label{eq_dust_optical_depth}\\
\chi       &= 71\, \left(\sigma_{d,21}/\mathcal{R}_{-16.5}\right)\,\left((G/G_{0})/(n/\cc)\right)\,,\label{eq_chi_full}
\end{align}
where $\Theta$ is the Heaviside function, $\tau_{\rm uv}$ the dust optical depth of the cloud, $\sigma_{d}^{-21}=\sigma_{d}/10^{-21}{\rm cm}^{-2}$ is the dust absorption cross section \citep{li_draine:2001apj}, $\mathcal{R}/10^{-16.5}{\rm cm}^{3}\,{\rm s}^{-1} $ is the formation rate coefficient of \HH~on dust grains \citep{wolfire:2008apj}, $G$ is the FUV flux in the Habing band ($6-13.6\,{\rm eV}$) normalized to the average Milky Way (MW) value $G_{0}$ \citep{habing:1968,draine:1978apjs}, and $n$ is the hydrogen number density. As in \citetalias{krumholz:2009apj}, we calculate the dust optical depth by linearly rescaling the MW value, i.e. $\tau_{\rm uv} = 10^{-21}{\rm cm}^{-2} N_{H}\, Z/\zsun /\mu$, where $N_{H}$ is the hydrogen column density and $\mu$ the mean molecular weight. In the simulation, the column density is calculated as $N_{H}= n\,l_{\rm cell}$; because of the mass threshold-based criterion used as a refinement in AMR, we expect $l_{\rm cell} \propto n^{-1/3}$, thus $N_{H} \propto n^{2/3}$.

Note that both $\sigma_{d}$ and $\mathcal{R}$ are proportional to the dust mass, that we assume to be proportional to the metallicity. Then the ratio between $\sigma_{d}$ and $\mathcal{R}$ is independent of $Z$. Additionally, eq. \ref{eq_fh2_full} can be simplified by assuming pressure equilibrium between the CNM and WNM. In this case, eq. \ref{eq_chi_full} turns out to be independent on $G/G_{0}$ and can be written as \citep{krumholz:2009apj}
\be\label{eq_sfr_last}
\chi = 0.75\,\left(1+3.1\,(Z/\zsun)^{0.365}\right)\,.
\ee 
\end{subequations}
As shown in \citep{krumholz:2011apj}, for $Z\gsim10^{-2}\zsun$ such approximation gives \HH~fractions compatible with those resulting from a full non-equilibrium radiative transfer calculations.

In Fig. \ref{fig_kmt_test} we plot $f_{\rm H2}$ from the \citetalias{krumholz:2009apj} model as a function of the gas density. Different solid lines refer to different metallicity. At a fixed metallicity, the molecular fraction as a function of density vanishes for low values of $n$; it steeply rises up to $f_{\rm H2} \sim 0.8$ in one density dex and asymptotically reaches $f_{\rm H2} = 1$. The critical density where the gas can be considered molecular ($f_{\rm H2}=0.5$) is roughly inversely proportional to the metallicity, i.e. $n_{c} \sim 25 (Z/\zsun)^{-1}\cc$ \citep[see also][]{agertz:2012arxiv}. We note that when detailed chemistry calculations are performed, such critical density depends on the chemical network and the assumptions regarding gas shielding from external radiation and clumpiness. As a consequence, the actual critical density can be higher that the one predicted by the \citetalias{krumholz:2009apj} model \citep[e.g.][]{bovino:2015arxiv}.

Because of the particular shape of the $f_{\rm H2}(n)$ relation, the adopted SF law (eqs. \ref{eq_sfr_tot}--\ref{eq_fh2_full}) is roughly equivalent to a prescription based on a density threshold criterion:
\begin{subequations}\label{eqs_sfr_equivalence}
\be
\dot{\rho}_{\star}=\Theta(n - n_{c}) m_p\,n /t_{\rm sf}\,,
\ee
where $m_p$ is the proton mass and the critical density
\be\label{eq_critical_density}
n_{c} \simeq 26.45 \, (Z/\zsun)^{-0.87} \cc\,
\ee
\end{subequations}
is calculated as a fit to the $f_{\rm H2}$ \citetalias{krumholz:2009apj} model. In Fig. \ref{fig_kmt_test}, we show $n_c$ for various metallicities (dashed vertical lines).

Eqs. \ref{eqs_sfr_equivalence} are not used to calculate the $SFR$ in the simulation. However, being simpler, such formulation can be used to enhance our physical intuition of the adopted SF law\footnote{As a consequence of the rough equivalence, it is not necessary to manually prevent SF in underdense regions, by imposing that an overdensity $\Delta>200$ is needed to form stars. At the start of the simulation ($z=100$), the mean density of the gas is $\sim 0.1\,m_p\,\cc$, while the \quotes{effective} SF threshold would be $n_{c} \sim 10^4\cc$ for gas at $Z=Z_{\rm floor}$.\label{footnote_sfr_equivalence}} in analyzing the results. As noted in \citet[][]{hopkins:2013arxiv}, the morphology of a galaxy is very sensitive to the minimum density of the cells that are able to form star.

During the simulation, eqs. \ref{eq_sfr_tot} are solved stochastically, by drawing the mass of the new star particles from a Poisson distribution \citep{rasera:2006,dubois:2008,pallottini:2014_sim}. We impose that no more than half of a cell mass can be converted into a star particle in each event. This prescription ensures the numerical stability of the code \citep{dubois:2008}. This is also consistent with the picture that nearly half of the mass in a MC is Jeans unstable \citep{federrath:2013}.

We allow SF only if the mass of a new star particle is at least equal to the baryon mass resolution. This avoids numerical errors for the star particle dynamics and enables us to treat the particle as a stellar population with a well sampled initial mass function (IMF). Additionally, the SF law is driven by \HH~formation on dust grains, we do not allow gas to form stars if the dust temperature is larger than $\simeq2\times 10^{3}$, because of dust sublimation (see Sec. \ref{sezione_blast} and App. \ref{app_rad_press} for the details on the dust prescriptions).

For the present work we assume a SF efficiency $\zeta_{\rm sf}=10\%$, in accordance with the average values inferred from MC observations \citep[][see also \citealt{agertz:2012arxiv}]{murray:2011apj}. Note that varying the parameters for the SF law should lead to similar $SFR$ once feedback are properly included, although the galaxy morphology can be different \citep{hopkins:2013arxiv}.

\subsection{Mass and energy inputs from stars}\label{sec_stellar_inputs}

Because of the finite mass resolution, it is necessary to introduce (according to eqs. \ref{eq_sfr_tot}--\ref{eq_sfr_last}) ``star particles'' to represent stellar populations. To this aim, we adopt a \citet{kroupa:2001} IMF
\begin{subequations}
\begin{align}
\Phi(m)\propto & \left[m^{-\alpha_{1}} \Theta(m_{1}-m)\right.\label{eq_imf}\\
+& \left. m^{-\alpha_{2}} \Theta(m-m_{1}) m_{1}^{\alpha_{2}-\alpha_{1}} \right]\,,\nonumber
\end{align}
where $\alpha_{1}= 1.3$, $\alpha_{2}= 2.3$, $m_{1} = 0.5\,\msun$, and $m$ is in the range $[10^{-1}-10^{2}]\msun$. The proportionality constant is chosen such that
%
\be
\int_{ 0.1\,\msun}^{100\,\msun} m\Phi\,{\rm d}m=1\, .
\ee
\end{subequations}

Once formed, stars affect the environment with chemical, mechanical and radiative feedback. These stellar inputs are parameterized by the cumulative fraction of the returned gas mass, metals and energy \citep[e.g.][]{salvadori:2008mnras,debennassuti2014mnras,salvadori:2015}. Mass and energy inputs are conveniently expressed per unit stellar mass formed ($M_{\star}$).

Chemical feedback depends on the return fraction ($R$) and the yield ($Y$):
\begin{subequations}\label{eqs_stellar_inputs}
\begin{align}\label{eqs_def_R_Y}
  R(t_{\star})	=&\int_{m(t_{\star}) }^{100\,\msun} (m-w) \Phi\,{\rm d}m\\
  Y(t_{\star})	=&\int_{m(t_{\star}) }^{100\,\msun} m_{Z} \Phi\,{\rm d}m\,,
\end{align}
where $w(m,Z_{\star})$ and $m_{Z}(m,Z_{\star})$ are the stellar remnant and the metal mass produced for a star of mass $m$ and metallicity $Z_{\star}$ \citep[e.g.][]{woosley:1995apjs,vandenhoek:1997a&as}, and $m(t_{\star})$ is the minimum stellar mass with lifetime\footnote{Stellar lifetimes are roughly independent of metallicity for $Z_{\star}>10^{-4}\zsun$ \citep[][see eq. 3]{raiteri:1996eq3}.} shorter than $t_{\star}$, the time elapsed from the creation of the stellar particle (i.e. the \quotes{burst age}).

This approach is used both in zoom galaxy simulations \citep[e.g.][]{agora:2013arxiv} and cosmological simulations \citep[e.g.][hereafter \citetalias{pallottini:2014_sim}]{pallottini:2014_sim}. Compared to cosmological simulations, though, zoom simulations have typically a better spatial and -- consequently -- time resolution (e.g. $\Delta t\sim 10^{-2}\,\myr$ vs $\Delta t\sim \myr$). Thus, here we can follow the gradual release of both gas and metals in the ISM.

The mechanical energy input includes SN explosions and winds, either by OB or AGB stars in young ($< 40\,\myr$) or evolved stellar populations:
\begin{align}\label{eqs_def_mec_energy}
  \epsilon_{\rm sn}(t_{\star}) =&\int_{m(t_{\star})>8\,\msun}^{40\,\msun }	e_{\rm sn}\Phi\,{\rm d}m,\\
  \epsilon_{\rm w}(t_{\star})  =&\int_{m(t_{\star})}^{100\,\msun }	 	e_{\rm w}\Phi\,{\rm d}m\,,
\end{align}
where $e_{\rm sn}=e_{\rm sn}(m,Z)$ and $e_{\rm w}=e_{\rm w}(m,Z)$ are the energy released by SN and stellar winds in units of $10^{51}{\rm erg}\equiv{\rm 1 foe}$; we have further assumed that only stars with $8 \leq m/\msun\leq40$ can explode as SN.

Radiative energy inputs can be treated within a similar formalism. The cumulative energy $\epsilon_{12}$ associated to the spectral range $(\lambda_{1}, \lambda_{2})$ can be written as
\begin{align}\label{eqs_def_rad_energy}
  \epsilon_{\rm 12}(t_{\star}) 	=&\int_0^{t_{\star}}\int_{m(t)}^{100\,\msun } L_{12}\Phi\,{\rm d}m\,{\rm d}t\\
  L_{\rm 12}(t) 		=& \int_{\lambda_{1}}^{\lambda_{2}}L_{\lambda}{\rm d}{\lambda}\,,
\end{align}
\end{subequations}
where $L_{\lambda}=L_{\lambda}(m,Z_{\star})$ is the luminosity per unit wavelength and mass. For convenience, we express the radiation energy in units of ${\rm foe}$, as for the mechanical energy (eqs. \ref{eqs_def_mec_energy}). In the following we specify $\epsilon_{\rm 12}$ in eq. \ref{eqs_def_rad_energy}, by separately considering ionizing radiation ($\lambda_{1}=0$, $\lambda_{2}=912\,\angstrom$) denoted by $\epsilon_{\rm ion}$, and the soft UV band, $\epsilon_{\rm uv}$, defined as the range ($\lambda_{1}=912\,\angstrom$, $\lambda_{2}=4000\,\angstrom$).

In eqs. \ref{eqs_stellar_inputs}, the quantities $w$, $m_{Z}$, $e_{\rm sn}$, $e_{\rm w}$, and $L_{\lambda}$ can be calculated from stellar evolutionary models. We adopt the {\tt padova} \citep{padova:1994} stellar tracks for metallicities $Z_{\star}/\zsun = 0.02,\, 0.2,\, 0.4,{\rm and}\, 1$ to compute the chemical\footnote{Similarly to \citet{agora:2013arxiv}, when computing the yields in eq. \ref{eqs_def_R_Y}, we assume that the metal mass is linked to the oxygen and iron masses via $m_{Z}= 2.09\,m_{\rm O} + 1.06\,m_{\rm Fe}$, as appropriate for \citet{asplund:2009ara&a} abundances.}, mechanical and radiative inputs using \textlcsc{starburst99} \citep{starburst99:1999,starburst99:2010apjs}.
\begin{figure}
\centering
\includegraphics[width=0.49\textwidth]{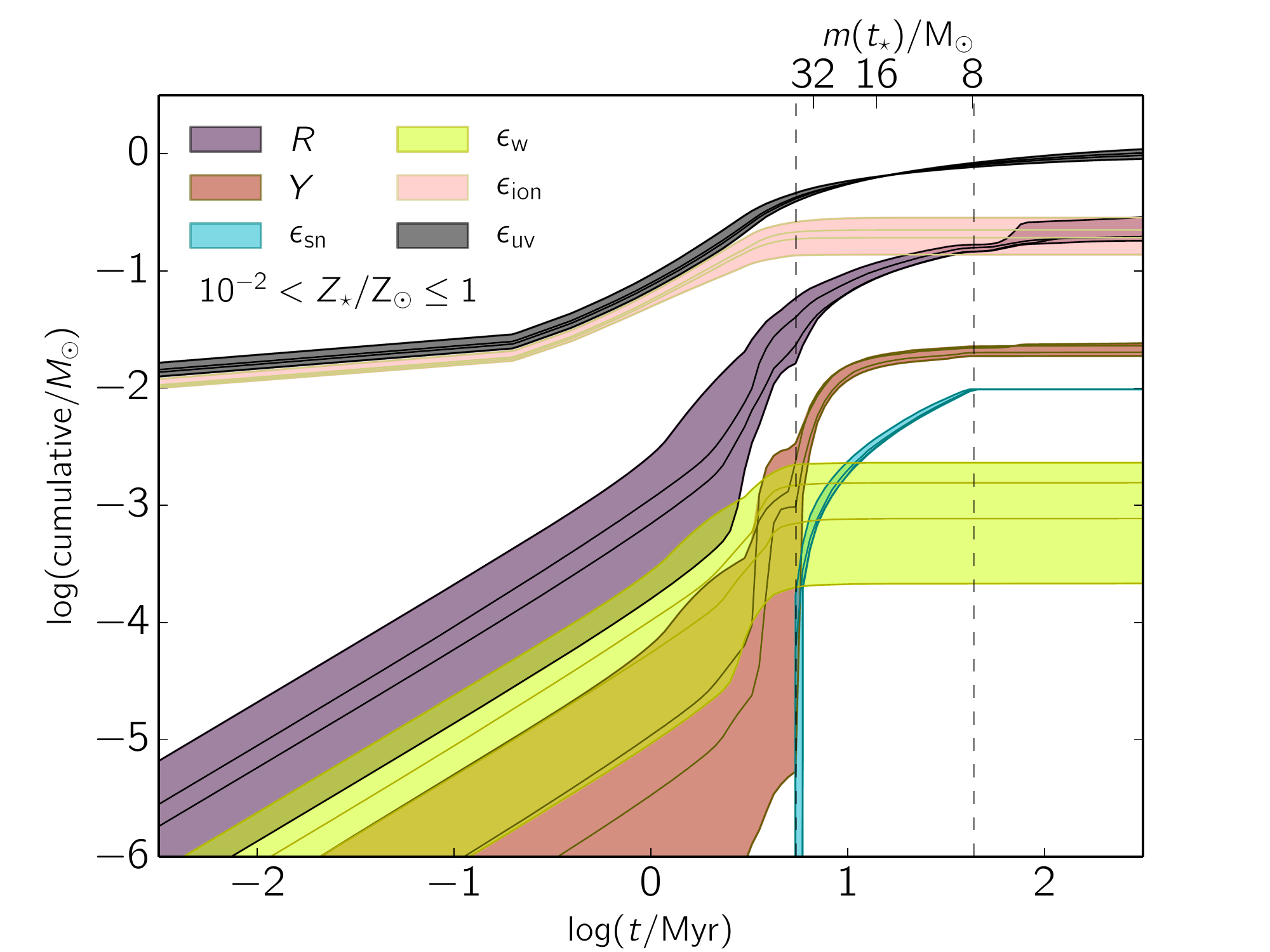}
\caption{
Stellar inputs (cumulative fraction) as a function of stellar age ($t_{\star}$). Shown are the return fraction ($R$), metal yield ($Y$), SN mechanical energy ($\epsilon_{\rm sn}$), wind mechanical energy ($\epsilon_{\rm w}$), ionizing radiation energy ($\epsilon_{\rm ion}$), and UV radiation energy ($\epsilon_{\rm uv}$).
The fractions are given per unit stellar mass formed; energies are expressed in units of $10^{51}{\rm erg}\equiv{\rm foe}$.
Cumulative fractions are indicated with a different colours, as indicated in the legend: the shaded regions cover the $0.02\leq Z_{\star}/\zsun\leq1$ metallicity range; dark lines denote single metallicity {\tt padova} stellar tracks \citep{padova:1994}.
To guide the eye, the SN explosion period is bracketed by vertical dashed lines; in the upper axis we report the value of $m(t_{\star})$, the minimum stellar mass corresponding to the stellar lifetime $t_{\star}$. For definitions, see eqs. \ref{eqs_stellar_inputs}.
\label{fig_gamete_tables}}
\end{figure}

In Fig. \ref{fig_gamete_tables} we plot $R$, $Y$, $\epsilon_{\rm sn}$, $\epsilon_{\rm w}$, $\epsilon_{\rm ion}$ and $\epsilon_{\rm uv}$ as a function of $t_{\star}$. For each curve the shaded regions denote the $0.02\leq Z_{\star}/\zsun\leq1$ metallicity range; single $Z_{\star}$ tracks are indicated with dark lines. The time interval during which massive stars can explode as SN ($0.8 \lsim \log t_{\star}/\myr\lsim 1.6$) is highlighted with vertical dashed lines, and the upper axis is labelled with the corresponding stellar mass.

Note that the OB stars contribution ($\log t_{\star}/\myr\lsim 0.8$) to $\epsilon_{\rm w}$, $Y$ and $R$ is roughly proportional to $t_{\star}$ and $Z_{\star}$ (see also \citealt{agertz:2012arxiv}, in particular eqs. 4). As in the simulation the metallicity floor is set to $Z_{\rm floor}=10^{-3}\zsun$, we slightly overestimate the wind contribution for low $Z_{\star}$.

Finally, note that the change of behavior of $\epsilon_{\rm uv}$ at $\log t_{\star}/\myr\lsim 2$ is due to the ionizing ($\lambda\leq\,912\,\angstrom$) photon production suppression. At late times ($\log t_{\star}/\myr\gsim 1.6$), AGB stars give a negligible mechanical energy contribution ($\epsilon_{\rm w}\simeq{\rm constant}$) but return mass and metals to the gas ($R$, $Y$).

\subsection{Stellar feedback}\label{sezione_blast}

\begin{figure}
\centering
\includegraphics[width=0.49\textwidth]{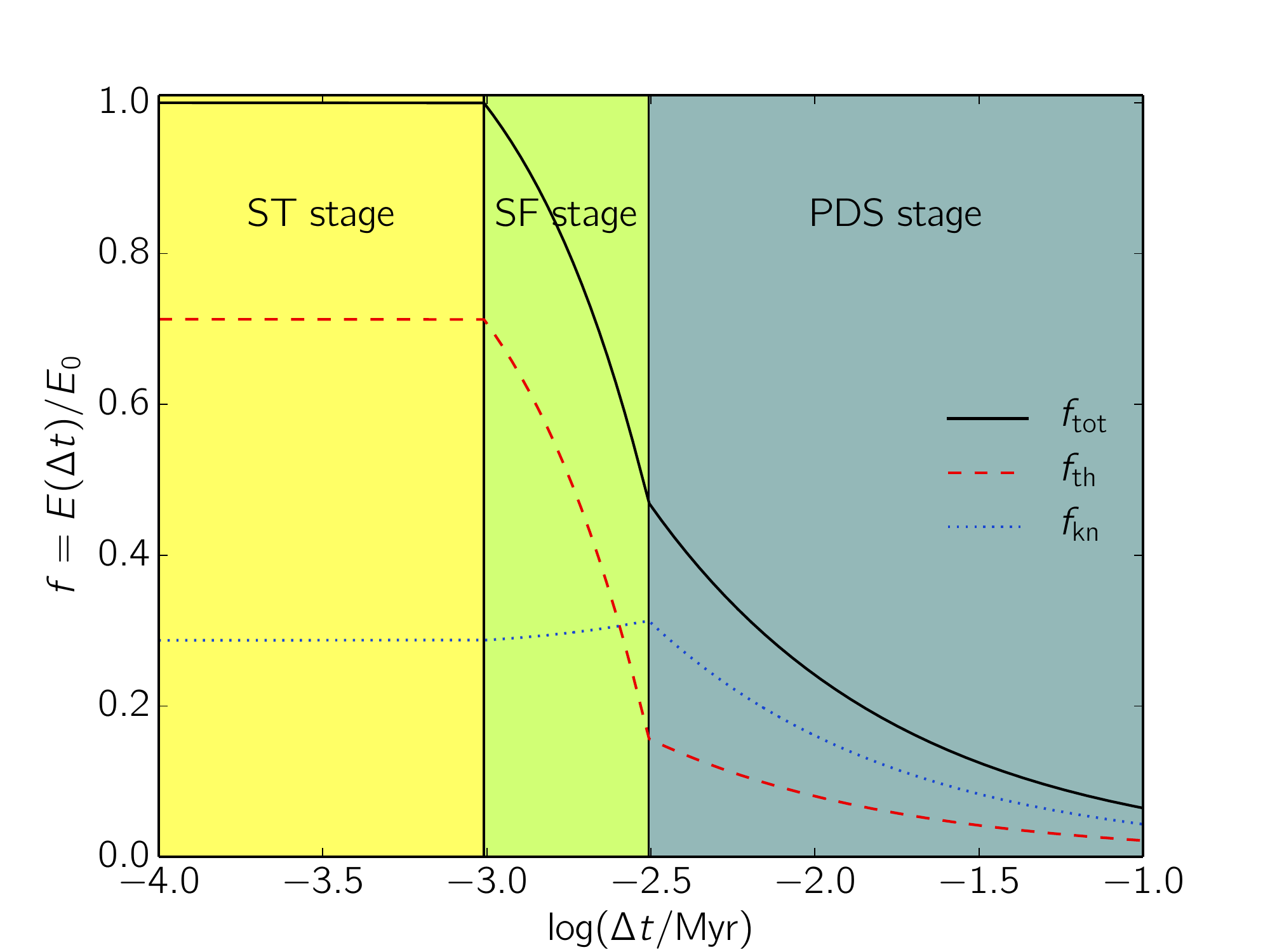}
\caption{
Example of the adopted feedback model. Fractional energy evolution for a single SN explosion ($E_{0}=1\,{\rm foe}$) in a gas characterized by $n=1\,\cc$ and $Z=10^{-3}\zsun$ as a function of the time interval from the explosion $\Delta t$. We plot the total ($f=f_{\rm th}+f_{\rm kn}$), thermal ($f_{\rm th}$), and kinetic ($f_{\rm th}$) energy fraction acquired by the gas (see eq. \ref{sn_energy_gas_equation}) with solid black, dashed red and dotted blue lines, respectively.
Shaded regions indicate different stages of the SN evolution, i.e. the energy conserving Sedov-Taylor (ST) stage, shell formation (SF) stage, and pressure driven snowplow (PDS).
In the adopted formalism, the initial energy $E_{0}$ is a function of the stellar input (see Sec. \ref{sec_stellar_inputs} and eq. \ref{eqs_def_mec_energy}), e.g. $E_{0}=[\epsilon_{\rm sn}(t_{\star}+\Delta t)-\epsilon_{\rm sn}(t_{\star})] M_{\star}$. The full model is presented in App. \ref{app_blastwave}.
\label{fig_blastwave_sketch}}
\end{figure}

Eqs. \ref{eqs_stellar_inputs} provide us with the energy produced by stars in different forms. The next step is to understand what fraction of that energy is eventually deposited in the ISM. Consider a stellar population of initial mass $M_{\star}$, metallicity $Z_{\star}$ and age $t_{\star}$ residing in a gas cell with volume $V_{\rm cell}$. In our scheme, when the simulation evolves for a time $\Delta t$, the chemical feedback act as follows:

\begin{subequations}
\begin{align}
  \rho 	&= \rho + \left[R(t_{\star}+\Delta t)-R(t_{\star})\right] M_{\star}/V_{\rm cell}\\
  Z    	&= Z 	+ \left[Y(t_{\star}+\Delta t)-Y(t_{\star})\right] M_{\star}/V_{\rm cell}\,,
\end{align}
where $\rho$ and $Z$ are the the gas density and metallicity and $R$ and $Y$ are taken from eqs. \ref{eqs_def_R_Y}. Note that chemical enrichment is due both to the SN and AGB winds.

\subsubsection{Supernova explosions}

For the mechanical feedback, let us first consider the case of SNe. At each SN event the specific energy of the gas changes as

\begin{align}\label{sn_energy_gas_equation}
 e_{\rm th} &= e_{\rm th}+ f_{\rm th} \left[\epsilon_{\rm sn}(t_{\star}+\Delta t)-\epsilon_{\rm sn}(t_{\star})\right] M_{\star}/V_{\rm cell}\\
 e_{\rm nth}&= e_{\rm nth}+ f_{\rm kn} \left[\epsilon_{\rm sn}(t_{\star}+\Delta t)-\epsilon_{\rm sn}(t_{\star})\right] M_{\star}/V_{\rm cell}\,,
\end{align}
\end{subequations}
where $e_{\rm th}$ and $e_{\rm nth}$ are the thermal and non-thermal energy densities, and $f_{\rm th}$ and $f_{\rm kn}$ are the fractions of thermal and kinetic energy deposited in the ISM. Thus, $e_{\rm nth}$ accounts for the momentum injection by SN and $e_{\rm th}$ for the thermal pressure part.

In the present work, we have developed a novel method to compute such quantities. The method derives $f_{\rm th}$ and $f_{\rm kn}$ from a detailed modelling of the subgrid blastwave evolution produced by the SN explosion. We calculate $f_{\rm th}$ and $f_{\rm kn}$ by evaluating the shock evolution at time $\Delta t$, the time step of the simulation\footnote{The underlying assumption is that the shock fronts exit the cell in $\lsim\Delta t$. This is quite consistent because the shock is expected to be supersonic, and the sound crossing time is larger or comparable with the simulation time step $\Delta t$, dictated by the Courant-Friedrichs-Lewy conditions.}.

The adopted blastwave model is based on \citet[][hereafter \citetalias{ostriker:1988rvmp}]{ostriker:1988rvmp}, and it accounts for the evolution of the blast through its different evolutionary stages (energy conserving, momentum conserving, etc.). While each stage is self-similar, the passage from one stage to the next is determined by the cooling time. Thus, $f_{\rm th}$ and $f_{\rm kn}$ depends on the blastwave evolutionary stage. The latter, in turn depends on the gas density, cooling time, and the initial energy of the blast ($E_{0}=[\epsilon_{\rm sn}(t_{\star}+\Delta t)-\epsilon_{\rm sn}(t_{\star})] M_{\star}$, in eq. \ref{eqs_def_mec_energy}).

The model details are presented in App. \ref{app_blastwave}. As an example, in Fig. \ref{fig_blastwave_sketch}, we show the energy evolution for a single SN explosion ($E_{0}=1\,{\rm foe}$) in a gas characterized by $n=1\,\cc$ and $Z=10^{-3}\zsun$. The total energy $E(t)$ is constant in the Sedov-Taylor (ST) stage, it decrease down to $0.5\,E_{0}$ during the shell formation (SF) stage, and it evolves as $\Delta t^{-2/7}$ in the pressure driven snowplow (PDS) stage (see eq. \ref{eq_energy_shock}). In the ST stage most of the energy is thermal, i.e. $f_{\rm kn}/f_{\rm th}\simeq 0.4$; however, in the SF stage $f_{\rm kn}$ increases, since part of the thermal energy is radiated away and some is converted into kinetic form\citep[e.g.][]{cox:1972apj,cioffi:1988apj}. Finally, during the PDS stage the ratio of thermal to kinetic is $f_{\rm kn}/f_{\rm th}\simeq 2$ (see eqs. 6.14 in \citetalias{ostriker:1988rvmp}).

In this particular example -- a $1\,{\rm foe}$ SN exploding in a $n=1\,\cc$ cell -- by assuming a simulation time step of $\Delta t\simeq 10^{-2}\myr$, we find that the blastwave is in the PDS stage, and the gas receives (via eqs. \ref{sn_energy_gas_equation}) a fraction of energy $f_{\rm th} \simeq 8\%$ and $f_{\rm kn} \simeq 16\%$ in thermal and kinetic form, respectively. During $\Delta t$, about $\simeq 75$\% of the initial SN energy has been either radiated away or lost to work done by the blastwave to exit the cell. The model is in broad agreement with other more specific numerical studies \citep[e.g.][]{cioffi:1988apj,walch:2015mnras,martizzi:2015mnras}.

\subsubsection{Stellar winds}

Stellar winds are implemented in a manner paralleling the above scheme for SNe. The energy variation can be calculated via eq. \ref{eqs_def_mec_energy}, where $\epsilon_{\rm sn}$ is substituted with $\epsilon_{\rm w}$, given in eqs. \ref{sn_energy_gas_equation}. Then, $f_{\rm th}$ and $f_{\rm kn}$ for winds are calculated via a stage scheme similar to SN. The main difference in the efficiency factors calculation depends on the mode of energy production, i.e. impulsive for SNe, continuous for winds. The complete scheme is detailed in App. \ref{app_blastwave}.

The efficiency of SN is greatly increased when the gas is pre-processed by stellar winds \citep{walch:2015mnras,fierlinger:2016}, since the energy loss process is highly non-linear \citep[][see Fig. 8]{fierlinger:2016}. For example, when a SN explodes in the lower density bubble produced by the stellar progenitor wind, the adiabatic phase lasts longer and consequently $f_{\rm kn}$ and $f_{\rm th}$ increase considerably. 

\subsubsection{Radiation pressure}\label{sec_rad_press}

Finally, we account for radiation pressure from stars. The coupling of the gas with the radiation can be expressed in terms of $\dot{p}_{rad}$, the rate of momentum injection \citep{krumholz:2009radpress,hopkins:2011mnras,krumholz:2012radpress,wise:2012radpres,agertz:2012arxiv}, and accounts for the contribution from ionization, and from dust UV heating and IR-trapping

\begin{subequations}
\begin{align}\label{eq_rad_moment_injection}
\dot{p}_{rad} =&  (L_{\rm ion}/c)(1-\exp(-\tau_{\rm ion})) \\
              +&   (L_{\rm uv}/c)((1-\exp(-\tau_{\rm uv})) +f_{\rm ir} )\,,\nonumber 
\end{align}
where $c$ is the speed of light, $\tau_{\rm ion}$ the hydrogen optical depth to ionizing radiation, and $f_{\rm ir}$ is the term accounting for the IR-trapping. $L_{\rm ion}$ and $L_{\rm uv}$ are calculated by integration of the stellar tracks (eqs \ref{eqs_def_rad_energy}). The calculation of $\tau_{\rm uv}$ is modelled in Sec. \ref{sec_model_sf} (eq. \ref{eq_dust_optical_depth} and related text). We compute $\tau_{\rm ion}$ and $f_{\rm ir}$ according to the physical properties of the gas, as detailed in App. \ref{app_rad_press}. Note that we do not assume, as sometimes done, $\tau_{\rm ion}\sim \tau_{\rm uv}\gg1$, i.e. we allow for the possibility that some LyC photons can escape.

In smoothed particle hydrodynamics (SPH) codes, radiation pressure (eq. \ref{eq_rad_moment_injection}) can be implemented as a \quotes{kick} \citep[e.g.][]{hopkins:2011mnras,barai:2015mnras}. Namely, a velocity $\Delta v = \dot{p}_{rad}\Delta t/m_b$ is directly added to some of the SPH particles of mass $m_b$ near the photon source. The particles that receive kicks are statistically chosen according to a probability $\mathcal{P}_{\rm kick}$, and with kick direction $\hat{v}$ that is sampled from a random distribution. Considering the specific kinetic energy of the SPH particles, we would have
\be
e_k = 0.5\,\left\langle m_b (\mathbf{v} + \Delta v \mathcal{P}_{\rm kick}\mathbf{\hat{v}} )^{2} \right\rangle/V_{cell}
\ee
where $\mathbf{v}$ is the original particle velocity, the $\langle\,\rangle$ operator indicates the particles sum weighted by the SPH kernel, and $V_{cell}$ is the kernel volume. Thus, because of the kick, the increase of energy density would be\footnote{In eq. \ref{eq_red_press_energy_increase}, when going from the first to the second line, note that first terms gives a null contribution, as $\mathbf{v}$ is ordered motion, while the kicks are randomly oriented via $\mathbf{\hat{v}}$, and that, by definition, $\langle \mathcal{P}_{\rm kick}\rangle = 1$.}

\begin{align}\label{eq_red_press_energy_increase}
\Delta e_k &= \langle m_b\, \mathcal{P}_{\rm kick} (\Delta v\mathbf{v}\mathbf{\hat{v}} + 0.5(\Delta v)^{2} \rangle/V_{cell} \nonumber\\
           &= 0.5\, m_b \, (\Delta v)^2/V_{cell} \\
           &= 0.5\,(\dot{p}_{rad}\Delta t)^2/(m_b\,V_{cell}) \nonumber\,,
\end{align}
\end{subequations}
where $\dot{p}_{rad}$ can be calculated via eq. \ref{eq_rad_moment_injection}, and eq. \ref{eq_red_press_energy_increase} can be directly cast into the AMR formalism. Additionally, because of our approximate treatment of IR-trapping (see App. \ref{app_rad_press}), we force energy conservation: $V_{cell} \Delta e_k \leq \Delta t\,(L_{\rm ion} + L_{\rm uv})$, i.e. the deposited energy must not exceed the radiative input energy. Finally, we recall here that non-thermal energy is dissipated with a time scale $t_{\rm diss}$, as described in the beginning of Sec. \ref{sec_numerical}.

\section{Results}\label{sec_result}

At $z=6$ ($t \simeq 920\, \myr$), the simulated zoom-in region contains a group of 15 DM haloes that host galaxies. We target the most massive halo ($M_{\rm h} = 1.8\times 10^{11}\msun$) that hosts \quotes{\emph{Dahlia}}, which is a galaxy characterized by a stellar mass of $M_{\star}=1.6\times 10^{10}\msun$, therefore representative of a typical LBG galaxy at that epoch. {Dahlia} has 14 satellites located within $\simeq 100 \,{\rm kpc}$ from its centre. The six largest ones have a DM mass in the range $M_{\rm h} = 2.5\times 10^{9}\msun - 1.2\times 10^{10}\msun$, and they host stars with total mass $M_{\star}\lsim 10^{9}\msun$. Additionally, there are eight smaller satellites ($M_{\rm h} \simeq 10^{7}\msun$), with $M_{\star}\simeq 10^{5}\msun$.

\subsection{Overview}\label{sec_res_barions}

\begin{figure*}
\centering
\includegraphics[width=0.99\textwidth,height=0.33\textwidth]{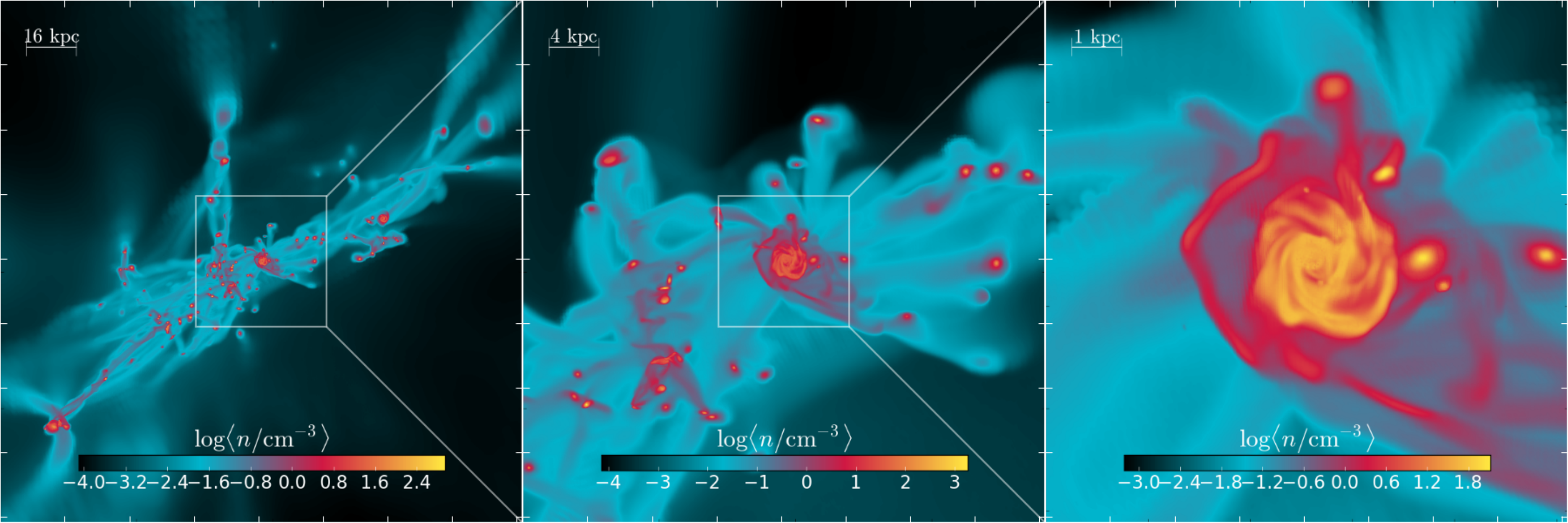}
\vspace{.3pt}

\includegraphics[width=0.99\textwidth,height=0.33\textwidth]{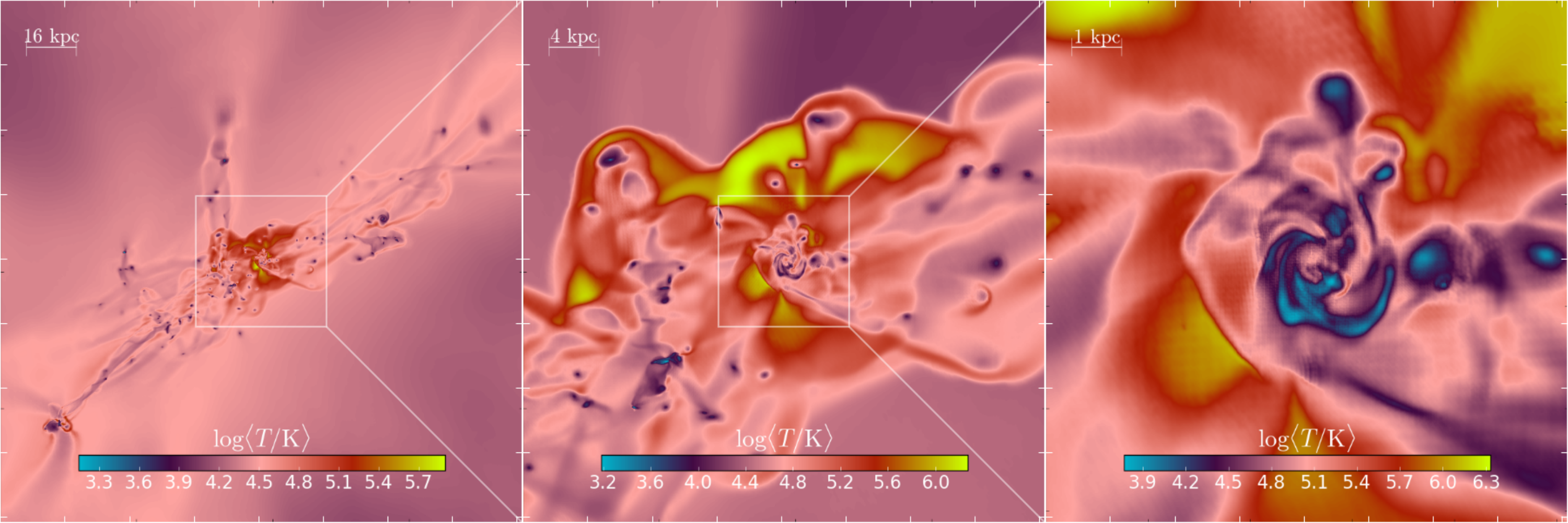}
\vspace{.3pt}

\includegraphics[width=0.99\textwidth,height=0.33\textwidth]{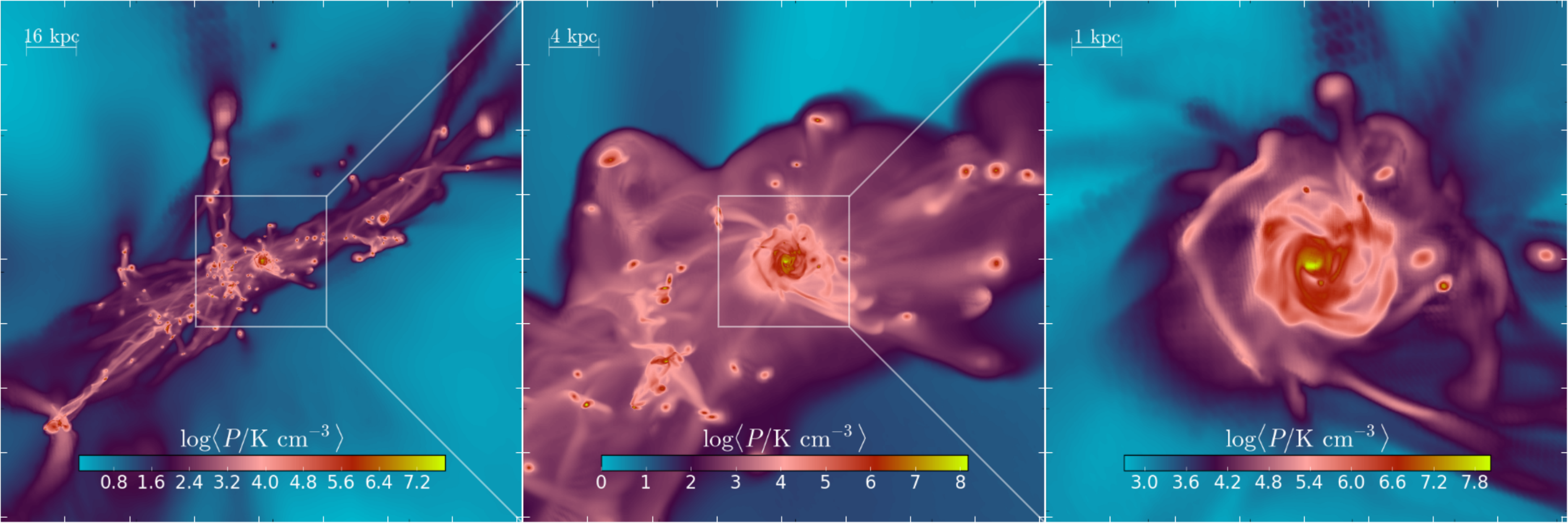}
\vspace{.3pt}

\includegraphics[width=0.99\textwidth,height=0.33\textwidth]{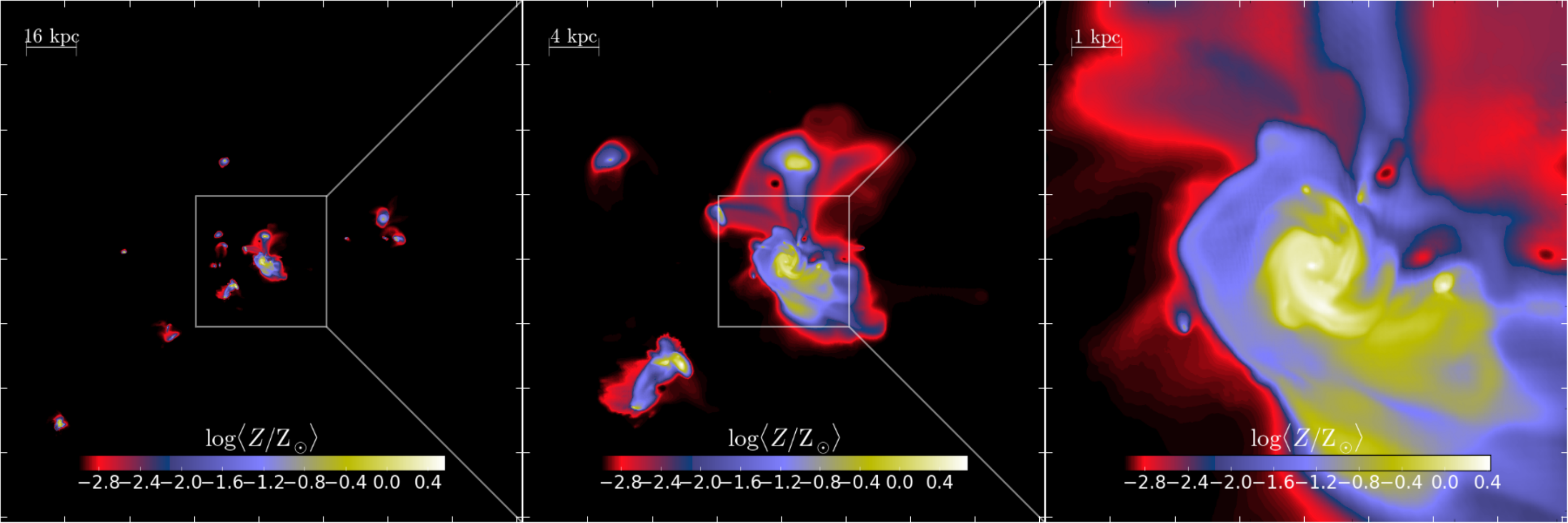}
\caption{
(Caption next page.) %
\label{fig_mappe_hydro}
}
\end{figure*}
\addtocounter{figure}{-1}
\begin{figure*}
\caption{(Previous page.) %
Maps of the simulated galaxy {Dahlia} $z=6$. From left to right we plot subsequent zooms on the galaxy. From top to bottom we plot the density ($n$), temperature ($T$), pressure ($P$) and metallicity ($Z$). Each map is obtained\textsuperscript{\ref{footnote_pymses}} by mass-averaging the physical quantity along the line of sight piercing the field of view and centred on {Dahlia}. In all panels the physical scale is indicated as an inset. Movies of {Dahlia} can be found at \url{https://www.researchgate.net/profile/Andrea_Pallottini}.
}
\end{figure*}

We start by looking at the overall properties of {Dahlia} on decreasing scales. In the following we refer to Fig. \ref{fig_mappe_hydro}, which shows the simulated density ($n$), temperature ($T$), total (thermal+kinetic) pressure ($P$), and metallicity ($Z$) maps\footnote{Most of the maps of this paper are obtained with a customized version of \textlcsc{pymses} \citep{labadens:2012aspc}, a visualization software that implements optimized techniques for the AMR grid of \textlcsc{ramses}.\label{footnote_pymses}}
at $z=6$.

\subsubsection{Environment (scale $\simeq 160$~kpc)}

{Dahlia} sits at the centre of a cosmic web knot and accretes mass from the intergalactic medium (IGM) mainly via 3 filaments of length $\simeq 100\,{\rm kpc}$, confirming previous findings \citep[][]{dekel:2009nat}. These overdense filaments ($n\simeq 10^{-2}\cc$) are slightly colder ($T\simeq10^{3.5}{\rm K}$) than the IGM ($\langle T \rangle\simeq10^{4.5}{\rm K}$) as a consequence of their shorter radiative cooling time ($t_{\rm cool}\propto n^{-1}$). Along these cold streams, pockets of shock-heated ($T\gsim10^{4.5}{\rm K}$) gas produced by both structure formation and feedback (SN and winds) are visible.

The galaxy locations can be pinpointed from the metallicity map, showing a dozen of metal-polluted regions. The size of the metal bubbles ranges from $\simeq20$ kpc in the case of {Dahlia} to a few ${\rm kpc}$ for the satellites. Bubble sizes increase with the total stellar mass (see \citetalias{pallottini:2014_sim}, in particular Fig. 13), and age of the galaxy stellar population. 

On these scales, the pressure is dominated by the thermal component ($P\simeq P_{\rm th}\sim 10^4{\rm K}\,\cc$); higher values of pressure, associated to non-thermal feedback effects (e.g. gas bulk motion), are confined around star forming regions, again traced by the metallicity distribution.

\subsubsection{Circumgalactic medium (scale $\simeq 50$~kpc)}\label{sec_CGM}

To investigate the circumgalactic medium (CGM), we zoom in a region within $\sim 3\, r_{\rm vir} = 47.5$ kpc from {Dahlia}'s centre. On these scales, we can appreciate the presence of several {Dahlia}'s satellites, i.e. extended (few ${\rm kpc}$) structures that are $\sim 100$ times denser than the filament in which they reside. Two of these density structures are particularly noticeable. These are located at a distance of $\sim10~{\rm kpc}$ from the centre in the upper left and lower left part of the map, respectively. By looking at the metallicity distribution, we find that both satellites reside within their own metal bubble, which is separated from Dahlia's one. This clearly indicates an in-situ star formation activity.

Additionally, the density map shows about $20$ smaller ($\sim 10-100\,{\rm pc}$) overdense clumps ($n\gsim 10\,\cc$). The ones within {Dahlia}'s metal bubble are enriched to $Z\simeq \zsun$. This high $Z$ value is indicative of in-situ self-pollution, which possibly follows an initial pre-enrichment phase from {Dahlia}. Clumps outside {Dahlia} metal bubble have on average an higher density ($n\sim 10^{2}\cc$). Since these clumps are unpolluted, they have not yet formed stars, as the effective density threshold for star formation is $\sim 25/(Z/\zsun)\cc$ (see eq. \ref{eq_critical_density} and Sec. \ref{sec_model_sf}). Such clumps represent molecular cloud complexes caught in the act of condensing as the gas streams through the CGM \citep{ceverino:2016MNRAS}. Such clumps have gas mass in the range $10^5 - 10^6 \msun$, and are not DM-confined, as the DM density field is flat on their location.

Star forming regions are surrounded by an envelope of hot ($T\simeq 10^{5.5}{\rm K}$), diffuse ($n\gsim 10^{-2}\,\cc$) and mildly enriched ($Z\sim 10^{-2}\zsun$) gas produced by SN explosions and winds. In the centre of star forming regions, instead, the gas can cool very rapidly due to the high densities/metallicities. Nevertheless, these regions are highly pressurized due to bulk motions mostly driven by radiation pressure (see Fig. \ref{fig_feedback_vs_time}).

\subsubsection{ISM (scale $ \simeq 10$~kpc)}\label{sec_small_scale}

The structure of Dahlia's ISM emerges once we zoom in a region $\sim 0.5\, r_{\rm vir}$ from its centre. In the inner region ($\simeq 2\,{\rm kpc}$), a counterclockwise disk spiral pattern is visible, since the field of view is perpendicular to the rotation plane of the galaxy (see \citealt{gallerani:2016outflow} for the analysis of the velocity field of {Dahlia}). The presence of disks in these early systems has already been suggested by other studies. For example, \citet{feng:2015apj} show that already at $z \sim 8$ nearly $70\%$ of galaxies with $M_{\star}\simeq 10^{10}\msun$ have disks (see also Sec. \ref{sec_final_results}).

The spiral central region and the spiral arms are dense ($n\simeq 10^{2}\cc$) and cold ($T\simeq 10^3{\rm K}$), and the active SF produces a large in-situ enrichment ($Z\simeq \zsun$). Winds and shocks from SN have no effect in the inner part of the galaxy, because of the high density and short cooling time of the gas; this implies that metals remain confined within $\sim 2\,{\rm kpc}$.
 Within spiral arms radiation pressure induced bulk motions largely dominate the total pressure, which reaches values as high as $P\gsim 10^{6.5}{\rm K}\,\cc$. The imprint of SN shocks is evident in the temperature map in regions with $T\gsim 10^5{\rm K}$. Shock driven outflows originated in spiral arms travel outward in the CGM, eventually reaching the IGM if outflow velocities exceed the escape velocity ($\sim 100\,{\rm km}\,{\rm s}^{-1}$, see Fig. 4 in \citealt{gallerani:2016outflow}).

Outflows are either preferentially aligned with the galaxy rotation axis, or they start at the edge of the disk. However, when spherically averaged, infall and outflow rates are nearly equal ($\sim 30\,\msun/{\rm yr}$ at $z\sim6$, \citealt{gallerani:2016outflow}), and the system seems to self-regulate \citep[see also][]{dekel:2014}.

Outside the disk, clumps with density $n\simeq 10^{2}\cc$ are also present and are actively producing stars. These isolated star forming MCs are located at a distance $\gsim 2\,{\rm kpc}$ from the centre, and show up as spots of high pressure ($P\gsim 10^7{\rm K}\,\cc$); some of this MCs are completely disrupted by internal feedback and they can be recognized by the low metallicity ($Z\sim 10^{-3}\zsun$): this is consistent with the outcome of numerical simulations of multiple SN explosions in single MC \citep[e.g.][]{kortgen:2016}.

\subsubsection{Radial profiles}

\begin{figure}
\centering
\includegraphics[width=0.485\textwidth]{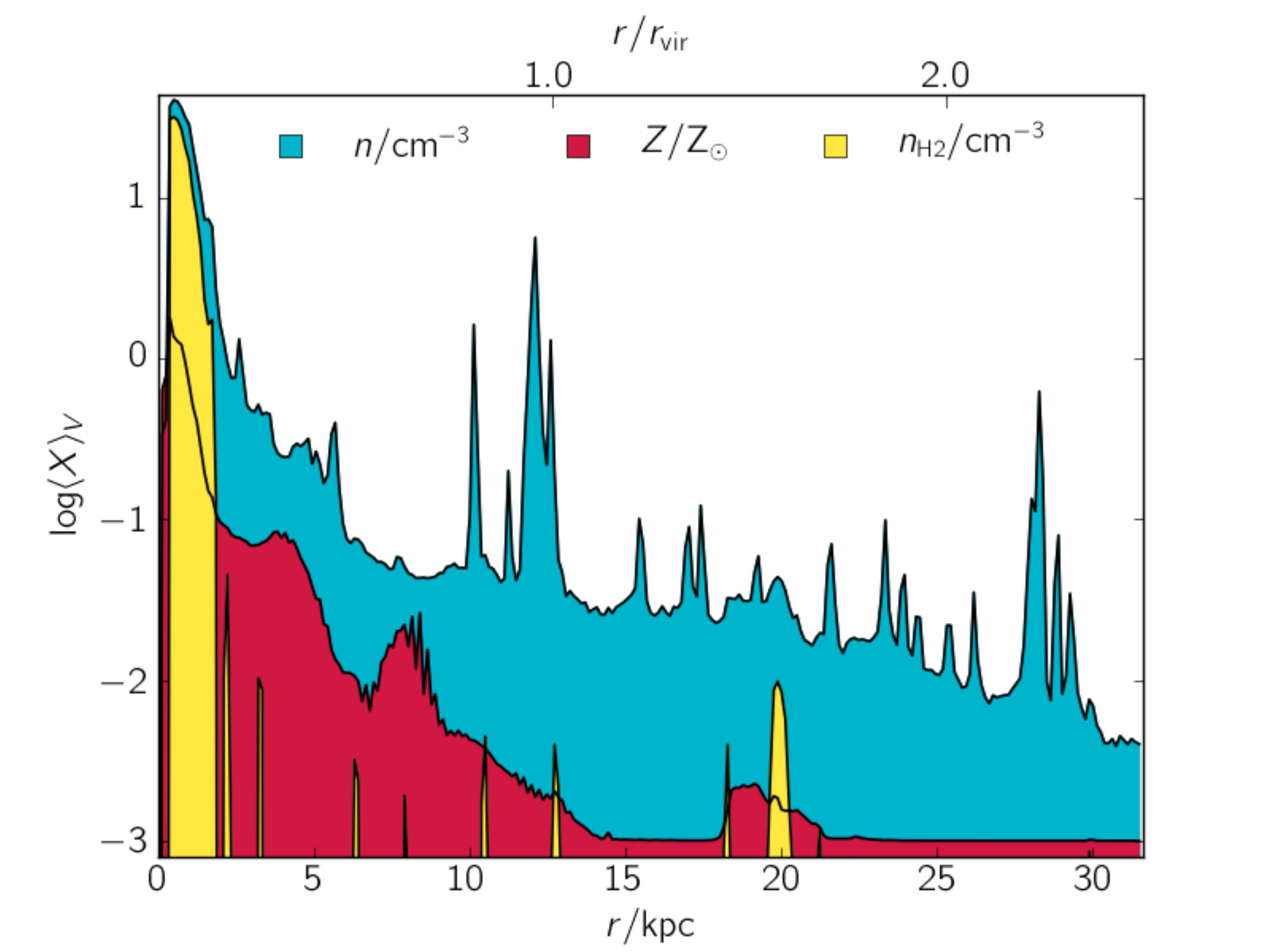}
\caption{
Density ($n$, blue), metallicity ($Z$, red) and molecular hydrogen density ($n_{\rm H2}$, yellow) radial profile ($r$) with respect to {Dahlia} centre. The profiles are spherically averaged, as indicated by the $\langle\,\rangle_{V}$ operator, and the upper axis shows the radial distance $r$ as a function of the virial radius of {Dahlia} ($r_{\rm vir}$).
\label{fig_sph_profile}
}
\end{figure}

Fig. \ref{fig_sph_profile} shows spherically averaged density, metallicity, and \HH~density profiles for the gas. The density profile rapidly decreases from $n\sim 30\,\cc$ at $r\sim 0$ to $n\sim 0.1\,\cc$ at $r\sim 6\,{\rm kpc} (\sim 0.5\,r_{\rm vir})$, and then flattens at larger distances. Such profile is consistent with the average profile of $z=4$ galaxies presented in \citetalias{pallottini:2014_sim}. There we claimed that the density profile is universal once rescaled to the halo virial radius (see also \citealt{liang:2016}). Superposed to the mean density profile, local peaks are clearly visible: they result from individual clumps/satellites, as discussed above. 

The central metallicity is close to the solar value, but by $r\sim12\,{\rm kpc}\sim r_{\rm vir}$ it has already dropped to $Z=Z_{floor}$. Within $0\lsim r/{\rm kpc}\lsim6$, the metallicity gradient closely tracks the density profile, while for $6\lsim r/{\rm kpc}\lsim15$ the decrease is steeper. \citet{pallottini:2014cgmh} find that the metallicity profile is not universal, however it usually extend up to few virial radii, as for {Dahlia}; further insights can be obtained by analyzing the $n$-$Z$ relation (Sec. \ref{sec_eos}).

In Fig. \ref{fig_sph_profile} we note that the $Z$ gradient found in {Dahlia} at $z = 6$ is slightly steeper than the one inferred from observations of $z\sim 3$ galaxies: i.e. we find $\Delta Z/r \sim -0.1\, {\rm dex}/{\rm kpc}$ while the observed ones are $\sim 0\, {\rm dex}/{\rm kpc}$ \citep{wuyts:2016} and $\sim +0.1\, {\rm dex}/{\rm kpc}$ \citep{troncoso:2013arxiv1311}. This suggests that the metallicity profile evolve with cosmic time and that the flattening is likely caused by stellar feedback, which in our Dahlia may occur in the following Gyr of the evolution. However, to prove such claim we should evolve the simulation to $z\sim3$.

The \HH~profile is spiky, and each peak marks the presence of a distinct SF region\footnote{We remind that the profiles are volume-weighted, thus the plotted $n_{\rm H2}$ accounts for the fact that \HH~is present only in a fraction of the gas at a given radius.}. In {Dahlia} \HH~is mainly concentrated within $r\lsim 0.5\,{\rm kpc}$ and it is distributed in the disk-like structure seen in Fig. \ref{fig_mappe_hydro} (see Sec. \ref{sec_final_results}). The location of the other peaks correspond to the satellites, which are mostly co-located with metallicity peaks. With increasing metallicity, in fact, lower densities are needed to form \HH~(eq. \ref{eq_critical_density}).

\subsection{Star formation and feedback history}\label{sec_sfr_result}

\begin{figure*}
\centering
\includegraphics[width=0.49\textwidth]{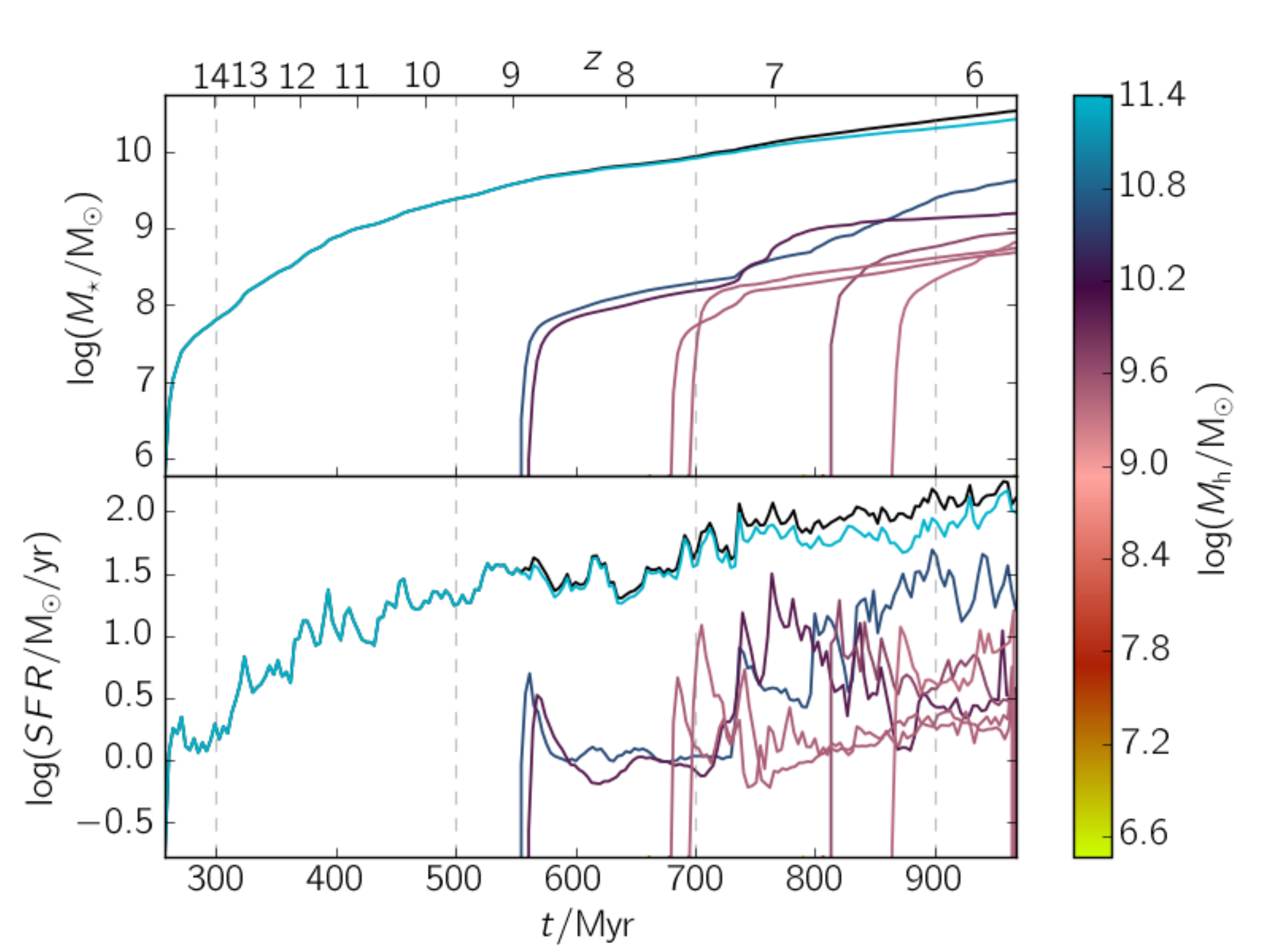}
\includegraphics[width=0.49\textwidth]{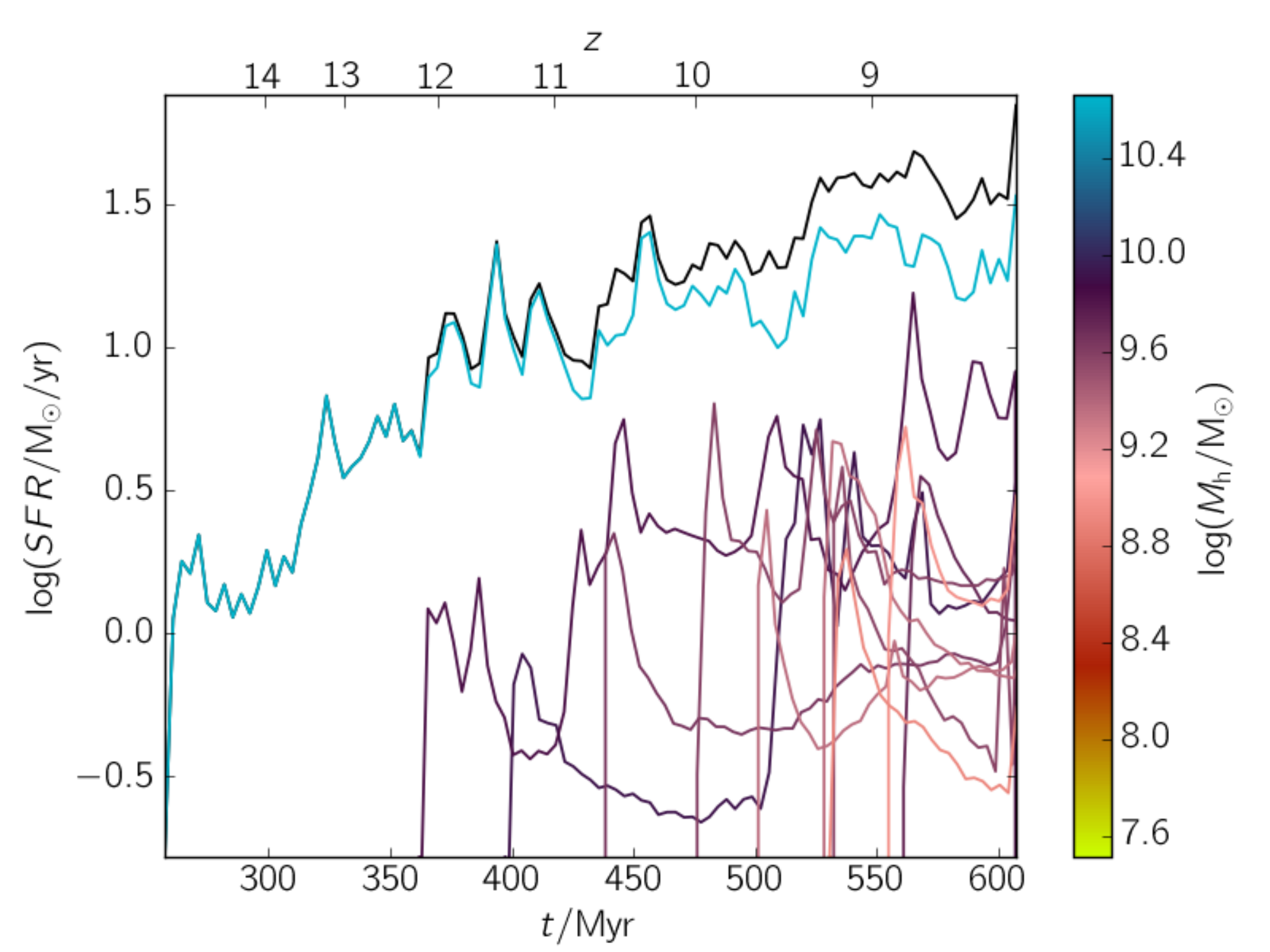}
\caption{
{\bf Left}: {Dahlia} and satellites cumulative stellar masses ($M_{\star}$, upper left panel) and star formation rates ($SFR$, lower left panel) as a function of cosmic time ($t$). For each galaxy, individual $M_{\star}$ and $SFR$ are plotted with a solid line, coloured accordingly to the total dark matter mass ($M_{\rm h}$) of the host halo at $z=6$. For both $M_{\star}$ and $SFR$, {Dahlia}'s tracks are plotted with a blue line, and the totals ({Dahlia}+satellites) are in black. {\bf Right:} $SFR$ as a function of cosmic time, with individual galaxies defined by the merger history up to $z\simeq8.5$. Note the different $M_{\rm h}$ colourbar scale with respect to the left panel. 
\label{fig_sfr_smf_energy}
}
\end{figure*}

We analyze the SF history of {Dahlia} and its major satellites by plotting in Fig. \ref{fig_sfr_smf_energy} the cumulative stellar mass ($M_{\star}$) and star formation rate ($SFR$) vs. time\footnote{The $SFR$ is averaged in steps of $\simeq 3\,\myr$. We have checked that smaller steps do not alter the following analysis.}.
%

For the whole galaxy sample, the time averaged ($\pm$ r.m.s.) specific star formation is $\langle{\rm sSFR}\rangle= (16.6 \pm 32.8)\,{\rm Gyr}^{-1}$. This mean value is comparable to that obtained by previous simulations of high-$z$ galaxies \citep{wise:2012radpres} and broadly in agreement with $z\sim 7$ observations \citep{Stark:2013ApJ}. At early times the $sSFR$ reaches a maximum of $\sim 100\,{\rm Gyr}^{-1}$, while a minimum of $3.0\,{\rm Gyr}^{-1}$ is found during the late time evolution. Both the large ${\rm sSFR}$ range and maximum at early times are consistent with simulations by \citet{shen:2014}. At late times, the $sSFR$ is in agreement with analytical calculation \citep{behroozi:2013apj}, and with $z=7$ observations \citep{gonzalez:2010}, although we note {Dahlia} has a larger stellar mass with respect to the galaxies in the sample ($M_{\star}\simeq 5\times 10^9\msun$).

At all times, {Dahlia} dominates both the stellar mass and star formation rate, whose mean value is $\langle SFR\rangle \simeq (35.3 \pm 32.7)\,\msun/{\rm yr}$. Its stellar mass grows rapidly, and it reaches $M_{\star}\sim 10^{9}\msun$ by $t\simeq 400\,{\rm Myr}$ ($z=11$), i.e. after $\simeq 120\,{\rm Myr}$ from the first star formation event. Such rapid mass build-up is due to merger-induced SF, that plays a major role at high-$z$ \citep{poole:2016MNRAS,behroozi:2013apj,salvadori2010MNRAS}. The $SFR$ is roughly constant from $z\sim 11$ to $z\sim 8.5$ and reaches a maximum of $\simeq 130\, \msun/{\rm yr}$ at $z\sim 6.7$. With respect to observations of $z\sim6$ LBG galaxies \citep[e.g.][]{stanway:2003MNRAS,stark:2009apj} the $SFR$ and $M_{\star}$ of {Dahlia} are above the mean values, but still consistent within one sigma. Additionally, the combination of $SFR$, $M_{\star}$, and $Z_{\star}$ for {Dahlia} are compatible with the fundamental mass metallicity relation observed in local galaxies \citep{mannucci:2010mnras}.

The total stellar mass in satellites is $M_{\star}\sim 10^{9}\msun$. Typically, SF starts with a burst, generating $\sim 10^{7.5} \msun$ of stars during the first $\simeq 20\,\myr$. Then the $SFR$ exponentially declines and becomes intermittent with a bursty duty cycle of $\sim100\,\myr$. This process can be explained as follows. As an halo forms, at its centre the density of the gas slowly rises. When the density is higher than the critical density of \HH~formation (eq. \ref{eq_critical_density}), the gas in the inner region is converted into stars in few free-fall times. Then feedback, and in particularly coherent SN explosions ($t_{\star}\gsim 10\,\myr$, see Fig. \ref{fig_gamete_tables}), quenches $SFR$, and the star formation activity becomes self-regulated. As mergers supply fresh gas, the $SFR$ suddenly goes out of equilibrium and becomes bursty again. Note that self-regulation is possible only for major satellites, since smaller ones ($M_{\rm h}\lsim 10^8\msun$) cannot retain a large fraction of their gas following feedback events due to their shallow potential wells (see \citetalias{pallottini:2014_sim}).

Note that the duty cycle and the amplitude of the burst are fairly in agreement with observations of $M_{\star}\sim10^8-10^{10}\msun$ galaxies at $z\lsim0.3$ \citep{kauffmann:2014mnras}. Furthermore, in our satellites we find that the typical behavior of the burst phases -- starburst - quiescent - post-starburst -- is qualitatively similar to what found by \citet{read:2016mnras}, that simulate the evolution of a $M_{\star}\simeq 10^9\msun$ galaxy for $\simeq 1\, {\rm Gyr}$ (see also \citealt[][]{teyssier:2013mnras,read:2016mnras_b} for further specific studies on the bursty nature of this kind of galaxies).

Since individual galaxies are defined as group of star particles in the same DM halo at $z=6$, the SF history accounts for the sum of all the stars that formed in different progenitors of the considered halo. For comparison, in the right panel of Fig. \ref{fig_sfr_smf_energy} we plot the $SFR$ of individual halos defined by their merger history at $z=8.7$. Galaxies with active SF at $300-550$ Myr merge into {Dahlia} at a later time, thus they do not appear individually in the left panel of Fig. \ref{fig_sfr_smf_energy}.

Superimposed to the global trend, the SF history of {Dahlia} and its satellites fluctuates on time scales of $\sim 10\,\myr$, corresponding to the time scale of energy deposition by feedback \citep[see e.g.][]{torrey:2016arxiv}.

\subsubsection{Star formation efficiency}\label{sez_sfr_efficiency}

\begin{figure}
\includegraphics[width=0.49\textwidth]{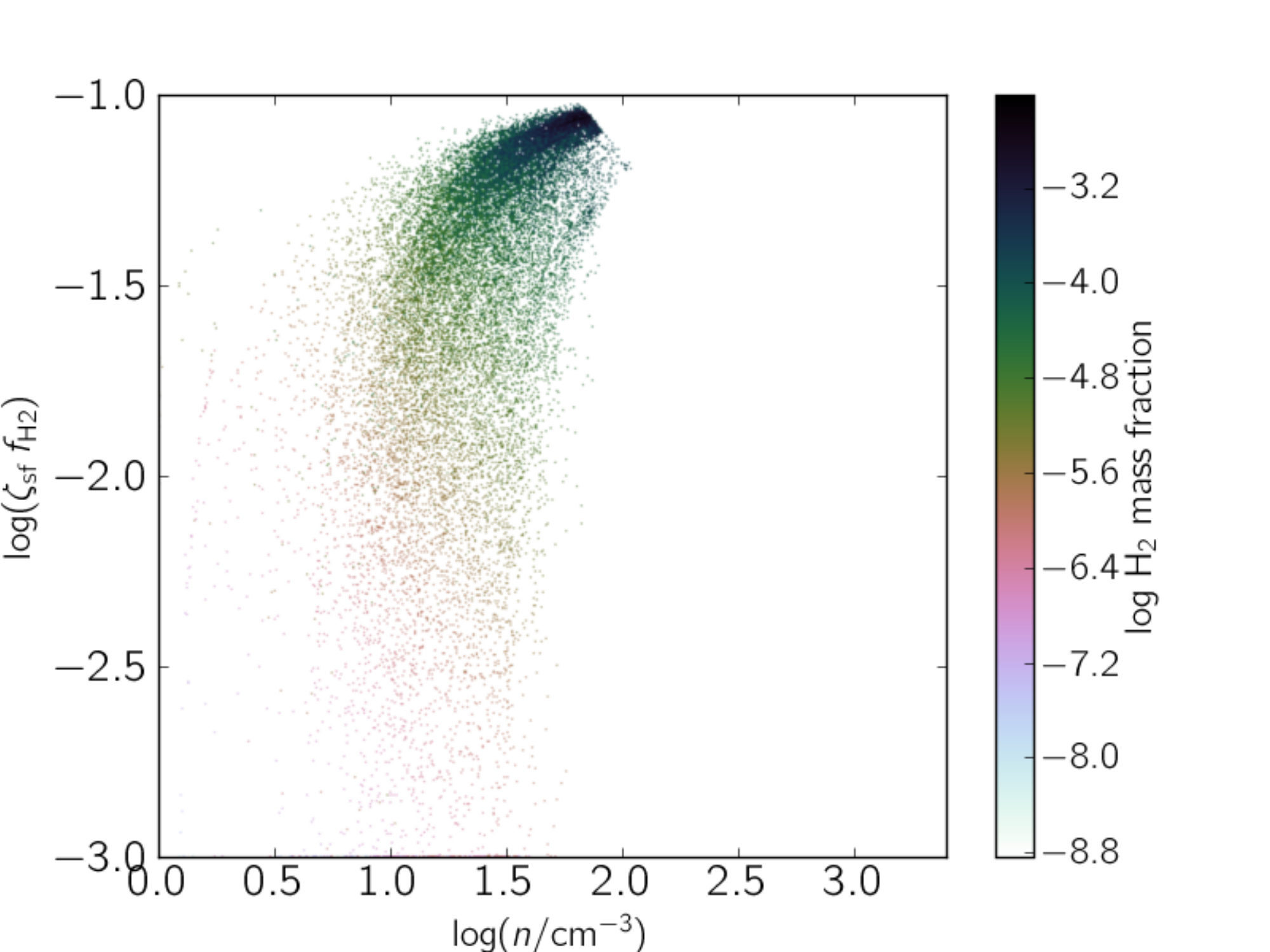}
\caption{Effective star formation efficiency ($\zeta_{\rm sf}\,f_{\rm H2}$) vs density ($n$) at $z=6$. The distribution is \HH~mass weighted; we consider gas within $3\, r_{\rm vir} = 47.5$ kpc from {Dahlia} centre.
\label{fig_cfr_semenov}}
\end{figure}

$\zeta_{\rm sf}\,f_{\rm H2}$ represents the quantity of gas converted in stars within a free-fall time (see eq. \ref{eq_sfr_tot}). In Fig. \ref{fig_cfr_semenov} we plot the effective star formation efficiency ($\zeta_{\rm sf}\,f_{\rm H2}$) as a function of gas density, weighted by the \HH~mass fraction at $z=6$. Most of the \HH~is contained in the range $n=10-100 \cc$, and the effective efficiency $\zeta_{\rm sf}\,f_{\rm H2}$ varies from $10^{-3}$ to $10^{-1}$. Since $\zeta_{\rm sf}= \mathrm{const.} =0.1$, the spread is purely due to the dependence of $f_{\rm H2}$ on density and metallicity (see Fig. \ref{fig_kmt_test}). Note that by construction $\zeta_{\rm sf}\,f_{\rm H2}\leq0.1$, and the plot does not show values very close to such limit, since gas with higher effective efficiency is converted into stars within a few free-fall times (eq. \ref{eq_sfr2}).

Interestingly, our \HH-based star formation criterion is reminiscent of a density threshold one, as below $n \simeq 3\, \cc$ the efficiency drops abruptly (eqs. \ref{eqs_sfr_equivalence}). However, an important difference remains, i.e. in the present model at any given density the efficiency varies considerably as a result of the metallicity dependence. The relation between efficiency and density is also similar to that found by \citet{semenov:2015} (\citetalias{semenov:2015}). This is striking because these authors use a star formation efficiency that depends on the turbulent velocity dispersion of the gas, with no notion of the local metallicity. This comparison is discussed further in Sec. \ref{sec_conclusioni}.

\subsubsection{Feedback energy deposition}\label{sec_feedback_res}

\begin{figure}
\centering
\includegraphics[width=0.49\textwidth]{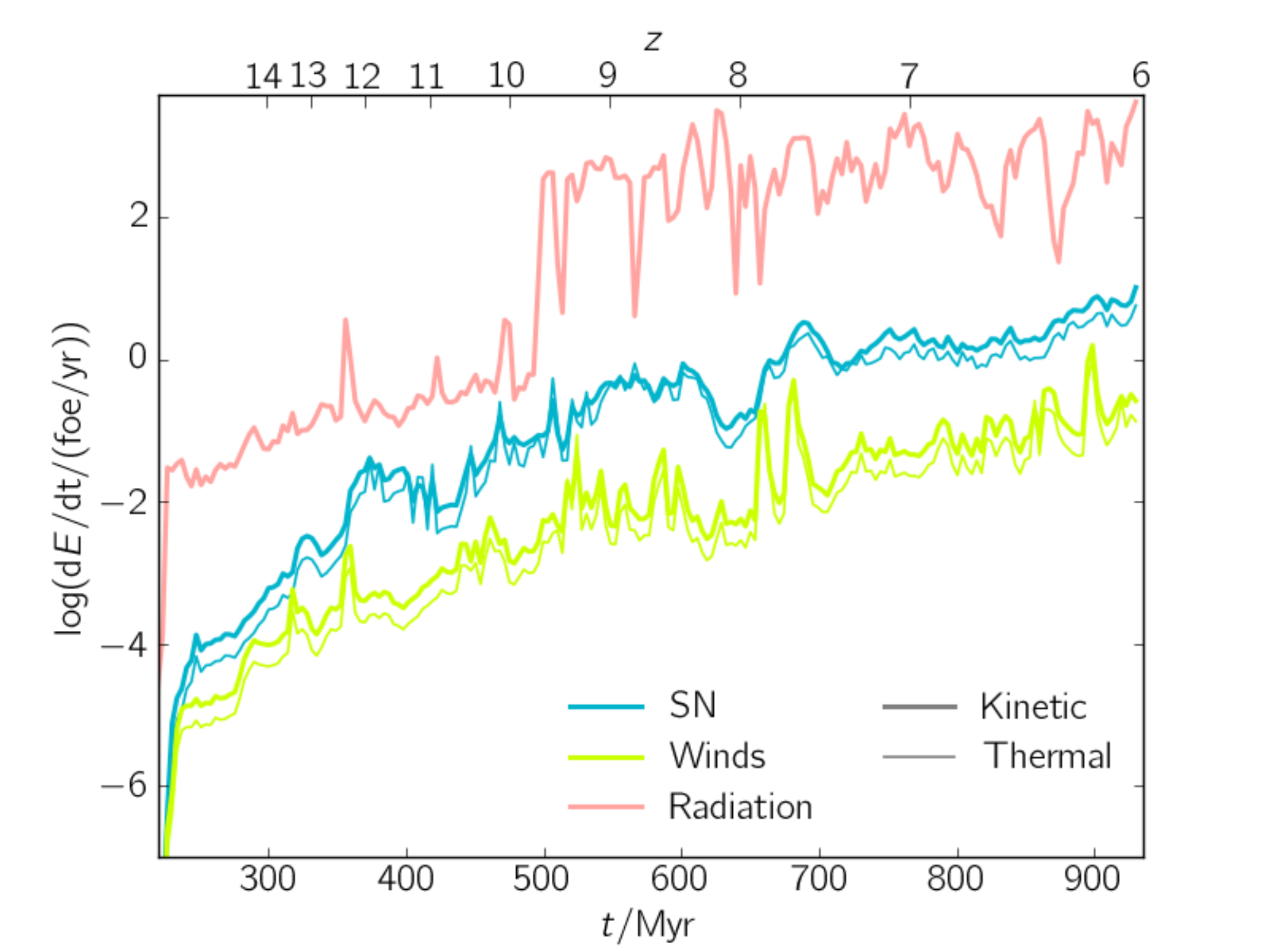}
\caption{Rate of energy deposition in the gas, ${\rm d}E/{\rm d}t$, by feedback processes as a function of cosmic time. Different contributions (SN, wind and radiation) are plotted with a different colour, and we additionally distinguish between the kinetic (thick lines) and thermal (thin lines) energy variation. By definition, radiation pressure has no thermal contribution. Note the jump at $t\simeq500\,\myr$ due to the onset of radiation pressure by AGB stars. The upper axis indicates the corresponding redshift.
\label{fig_feedback_vs_time}
}
\end{figure}

As discussed in Sec. \ref{sezione_blast}, only a small fraction of the available energy produced by stars can couple to the gas. During the simulation, we find that the time average efficiency of the conversion is $f \sim 0.1\%$, regardless of the feedback type. These low efficiencies imply that energy is mostly dissipated within MCs where the stars reside and produce it. For SN and winds, such small efficiency is a consequence of the short cooling times in MCs (see also App. \ref{app_blastwave}). For radiation pressure the efficiency is limited by the relatively small dust optical depths (see also App. \ref{app_rad_press}).

Note that, typically in simulations \citep[e.g.][]{wise:2012radpres,agertz:2012arxiv}, energy from stars is directly deposited in the gas, and then dissipation (mostly by radiative losses) occurs during the hydrodynamical time step. Within our scheme, instead, the deposited energy is already dissipated within high density cells, where cooling is important. Nevertheless, this does not appear to determine major differences in, e.g., $SFR$ history and ISM thermodynamics, as discussed in Sec. \ref{sec_eos}.

In Fig. \ref{fig_feedback_vs_time} we plot the energy deposition rate in the gas by various feedback processes as a function of time. Most evidently, \emph{radiation dominates the energy budget at all times}: $\dot E_{rad} \simeq 10^{2} \dot E_{SN}\simeq 10^3 \dot E_{w}$. The ratios of these energy rates somewhat reflect the stellar inputs shown in Fig. \ref{eqs_stellar_inputs}, although this is not a trivial finding, given that the interplay among different feedback types is a highly non-linear process.

As expected, the energy deposition rate behaves as $\dot E \propto SRF^q$, with $q \simgt 1$, apart from fluctuations and jumps as the one at $t\simeq 500\,\myr$. The scaling can be understood by simple dimensional arguments. Assume that most of the energy is deposited by radiation pressure. In the optically thick limit, we can combine eqs. \ref{eq_red_press_energy_increase} and \ref{eq_rad_moment_injection} to write $\dot E_{rad} \Delta t \simeq (L_{\rm uv} \Delta t)^2 / (M_{g}\,c^2)$, where $M_{g}$ is the gas mass accelerated by radiation, and we neglect ionizing radiation. Then, using \ref{eqs_def_rad_energy}, we can write $\dot E_{rad} \propto SFR\,(M_{\star}/M_{g})$. Initially, $M_{g}\simeq M_{\star}$, thus $\dot E_{rad} \propto SFR$. Once the gas mass is expelled from the star forming region or converted into stars, $M_{g}\ll M_{\star}$. Thus the deposition rate increases faster than the $SFR$ and it is very sensitive to the amount of gas mass around the sources.

The previous argument holds until the gas remains optically thick. This is warranted by AGB metal/dust production which becomes important after for stellar ages $t_{\star}\sim 100\,\myr$ (see Fig. \ref{fig_gamete_tables}). When combined with the parallel increase of UV photons by the same sources, it is easy to interpret the rapid increase of the radiative feedback efficiency at $t\simeq500\,\myr$, i.e. after $\simeq 200 \,\myr$ from the first star formation events in {Dahlia}. We checked this interpretation by looking at the IR-trapping recorded on the fly during the simulation. We find that on average $f_{\rm ir}\simeq 10^{-2}$ for $t\lsim 500\, \myr$, and $f_{\rm ir}\simeq 0.1$ at later times, thus confirming our hypothesis.

The energy deposition rates for different feedback types are highly correlated in time (Pearson coefficients $\gsim 0.7$). This is partially due to the fact that the same stellar population inputs wind, radiation and supernova energy in the gas. Additionally, as we have just seen for the case of AGB star, different types of feedback are mutually dependent. For example, radiation pressure is more effective when the gas is metal and dust enriched by SN and AGB stars; winds and SN can more efficiently couple with low density gas (longer cooling time).

Note that short and intense peaks in energy deposition rate correspond to the complete disruption of multiple MCs. This occurs following strong SF events in small satellites ($M_{\rm h} \sim 10^{7}\msun$) that cannot retain the gas and sustain a continuous star formation activity.

Finally, we remind that, when compared with observational/analytical constraints, the $SFR$ and $M_{\star}$ of {Dahlia} are higher then the mean, but still consistent within one sigma. We caution that this might imply a somewhat weak feedback prescription.

\subsubsection{Feedback effects on ISM thermodynamics}\label{sec_eos}

\begin{figure*}
\centering
\includegraphics[width=0.485\textwidth]{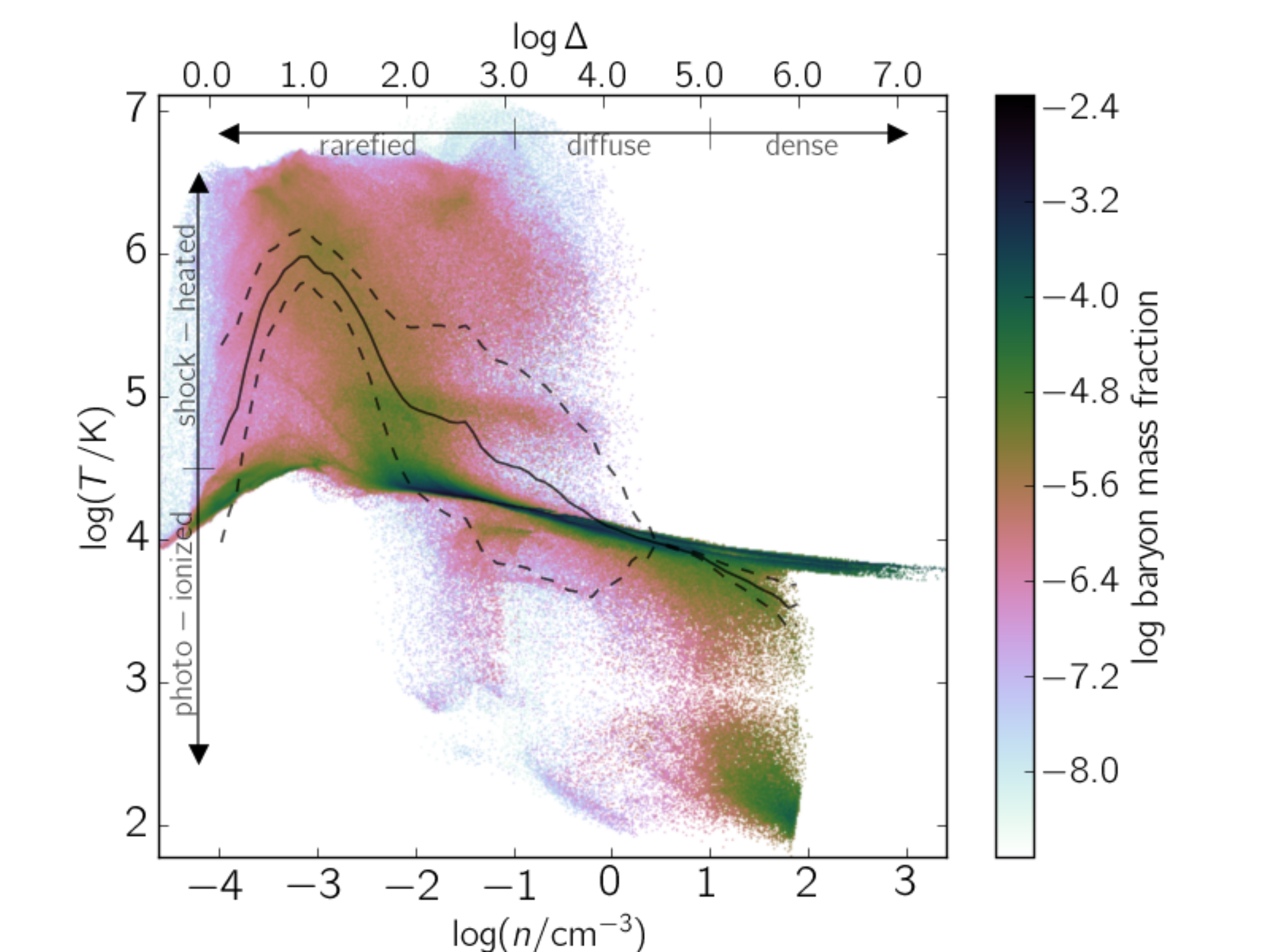}
\includegraphics[width=0.485\textwidth]{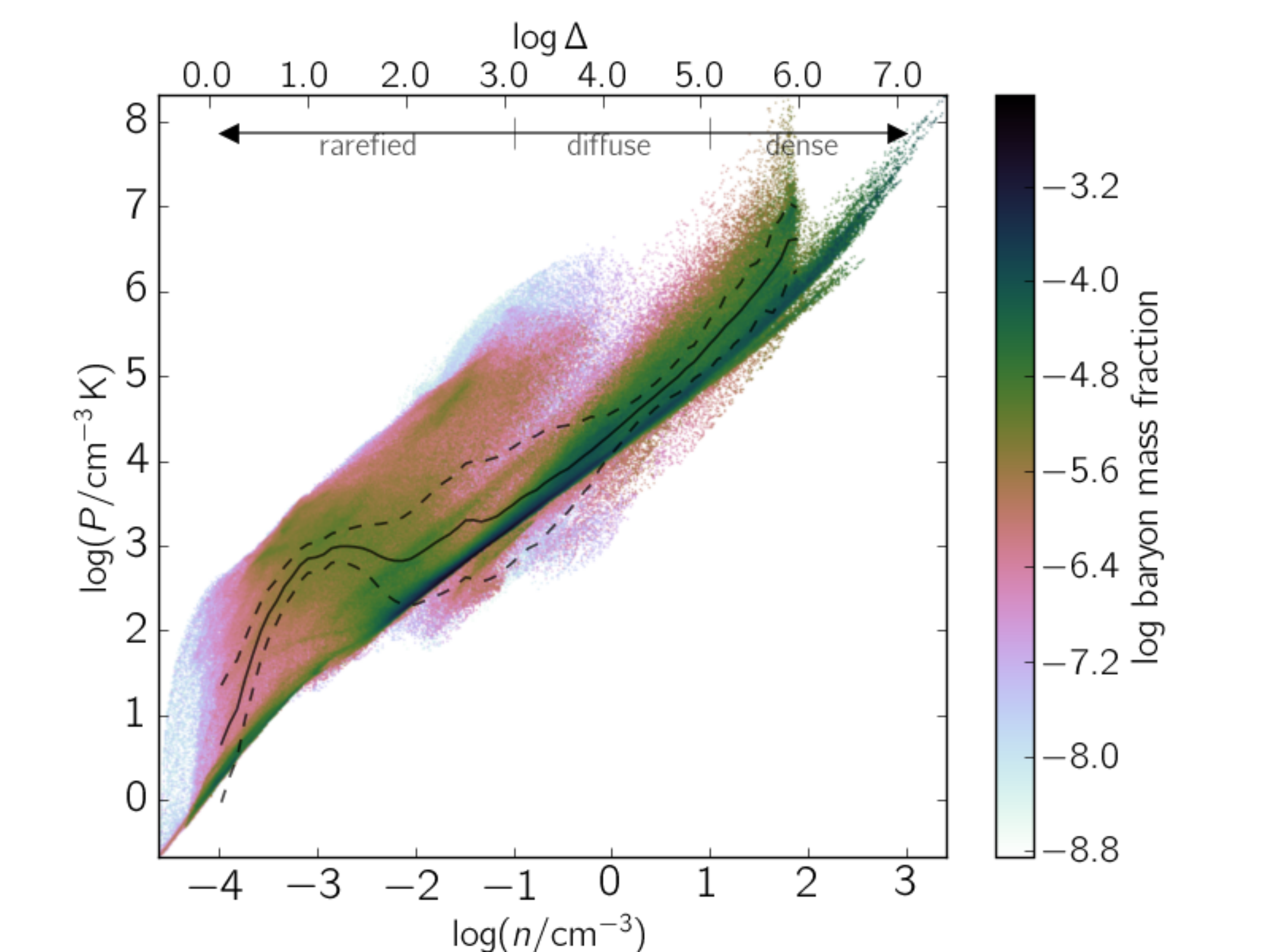}

\includegraphics[width=0.485\textwidth]{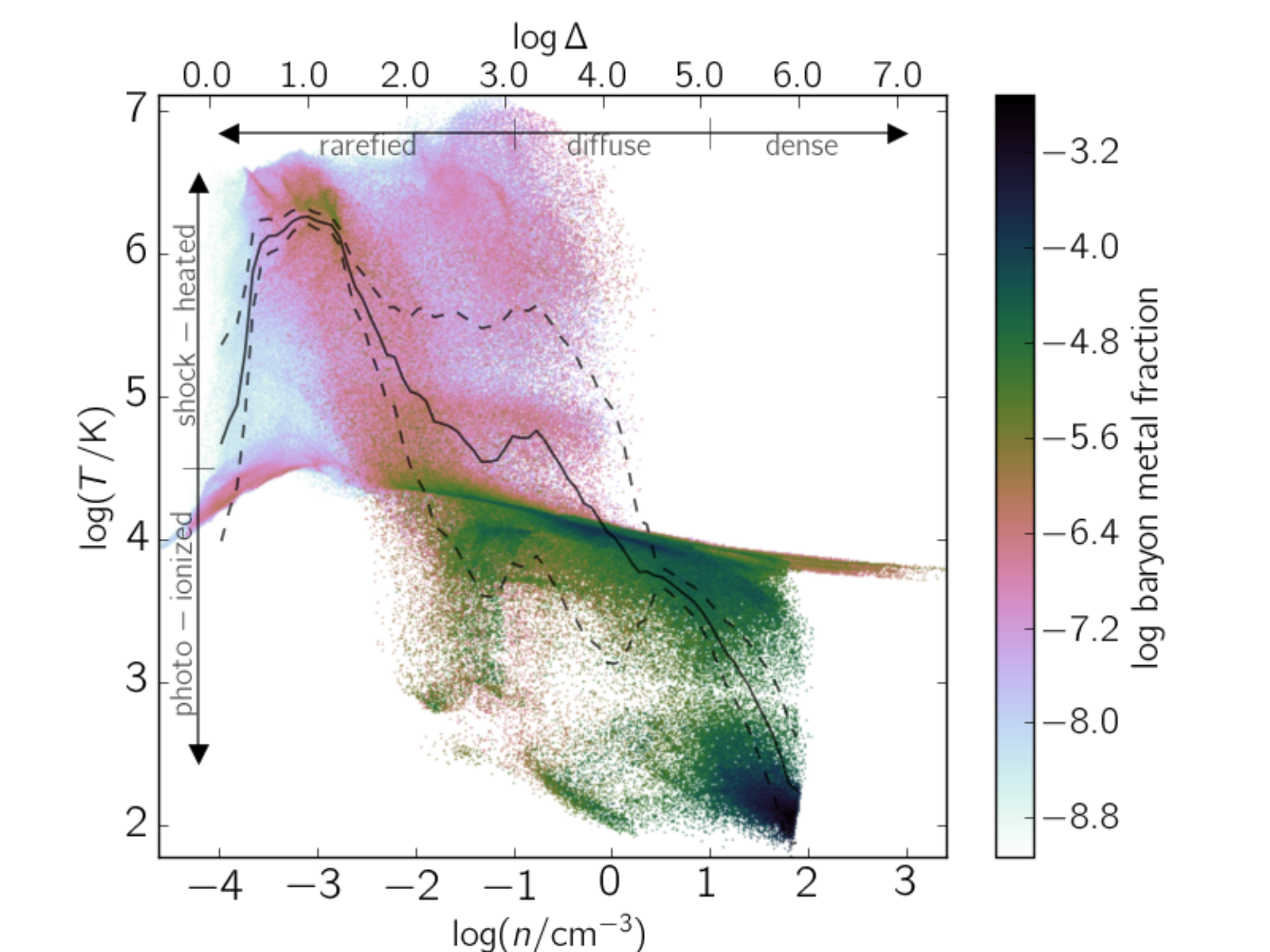}
\includegraphics[width=0.485\textwidth]{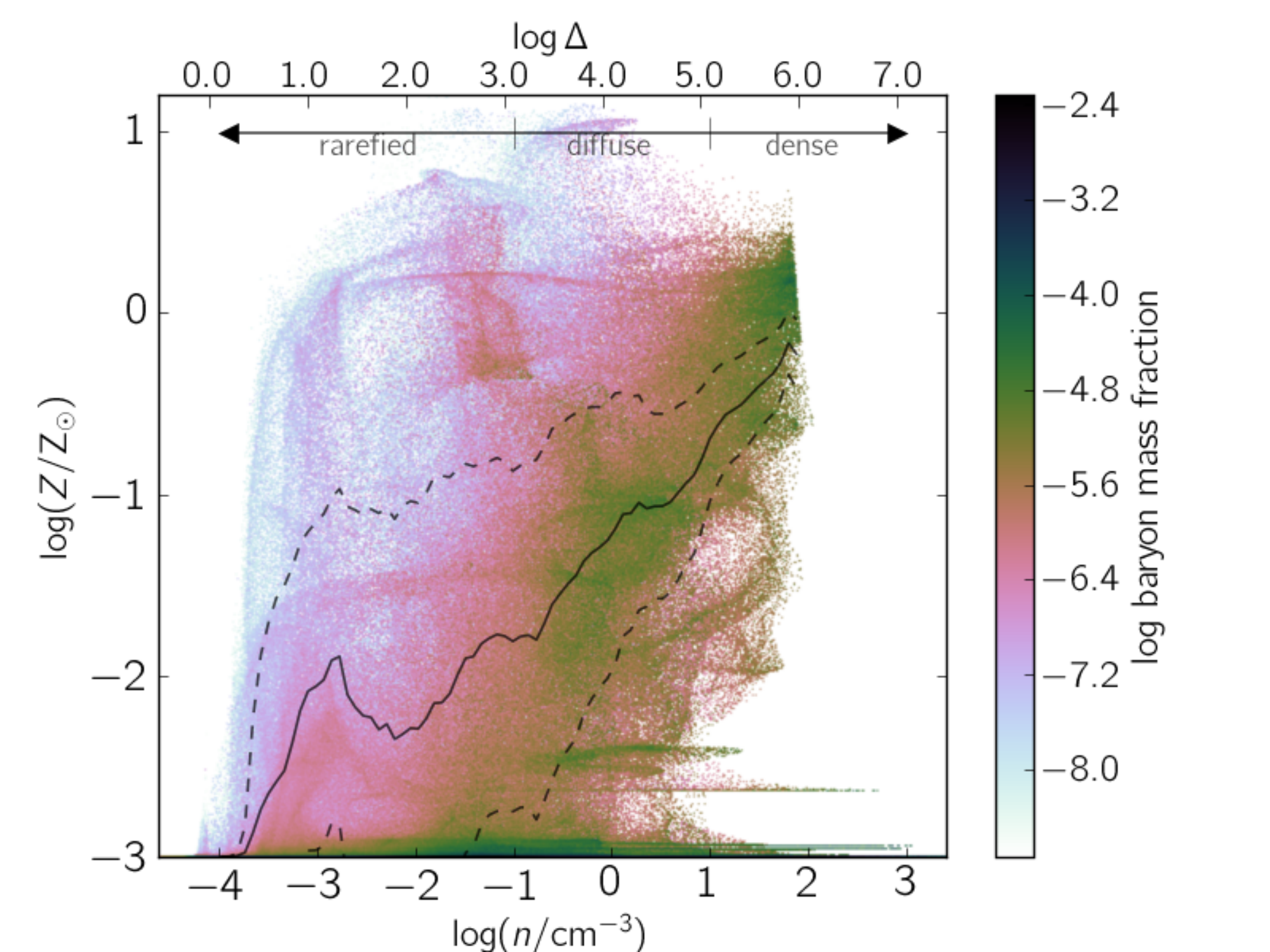}
\caption{
Equation of State of the gas within $\simeq47.5\,{\rm kpc}$ $(3\, r_{\rm vir})$ from {Dahlia} centre at $z=6$. Each EOS consists in a mass- or metal-weighted probability distribution function (PDF) as specified by the colourbar. We plot the PDF in the $n$-$T$ plane ({\bf upper left panel}), in the $n$-$P$ plane ({\bf upper right panel}), the metal-mass weighted PDF in the $n$-$T$ plane ({\bf lower left panel}), and mass-weighted relation between gas $n$ and $Z$ ({\bf lower left right}). Mean relations and r.m.s. dispersions are overplotted with solid black and dashed lines, respectively. In the upper horizontal axis of each panel, we indicate the overdensity ($\Delta$) corresponding to $n$. The density range of rarefied, diffuse and dense phases used in the text are indicated. For the panels on the left, the rarefied gas is additionally divided in \emph{photo-ionized} ($T<10^{4.5}{\rm K}$) and \emph{shock-heated} ($T\geq10^{4.5}{\rm K}$). See Tab. \ref{tagella_eos_riassunto} for a summary of the total values.
\label{fig_eos_1}
}
\end{figure*}

\begin{table}
\centering
\begin{tabular}{lcccc}
\hline\hline
& mass & rarefied & diffuse & dense \\
\hline
Gas & $1.3 \times 10^{10}\msun$ & $44\%$ & $34\%$ & $22\%$ \\
Metals & $ 4.2\times 10^5\msun$ & $5\%$ & $25\%$ & $70\%$\\
\HH & $ 3.6\times 10^8\msun$ & $0\%$ & $1\%$ & $99\%$\\
\CIIion & $ 2.2\times 10^5\msun$ & $4\%$ & $22\%$ & $74\%$\\
\end{tabular}
\caption{
Summary of the gas masses for total, metal, \CIIion, and \HH~within $\sim 47.5\,{\rm kpc}$ $(3\, r_{\rm vir})$ from {Dahlia} center. In the table, we report also the fraction that is contained in different gas phases\textsuperscript{\ref{footnote_phases}}: \emph{rarefied} ($\log(n/{\rm cm}^3)\leq -1$), \emph{diffuse} ($-1<\log(n/{\rm cm}^3)\leq 1$) and \emph{dense} ($\log(n/{\rm cm}^3)> 1$). Discussion about gas and metal mass is found in Sec. \ref{sec_eos}; analysis of \HH~and \CIIion~is in Sec. \ref{sec_final_results} (see also App. \ref{sez_cloudy_model} for \CIIion~calculation).
\label{tagella_eos_riassunto}}
\end{table}

Feedback leaves clear imprints in the ISM thermodynamics. For convenience, we classify ISM phases according to their density: we define the gas to be in the \emph{rarefied}, \emph{diffuse}, and \emph{dense} phase if $n \leq 0.1\,\cc$, $0.1 \leq n/\cc\leq 10$, $n > 10\,\cc$, respectively\footnote{Compared to the definitions used in \citet{klessen:2014review}, the rarefied corresponds to the warm and hot ionized medium, the diffuse phase to the cold and warm neutral medium and the dense phase to the molecular gas.\label{footnote_phases}}.

We focus at $z=6$ and consider the gas in a region within $\simeq 47.5\,{\rm kpc}$ $(3\, r_{\rm vir})$ from {Dahlia}'s centre, essentially the scale of the CGM described in Sec. \ref{sec_CGM}. This region contains a total gas mass of $1.3\times 10^{10} \msun$, and metal mass of $4.2\times 10^5\msun$ (additional data in Tab. \ref{tagella_eos_riassunto}).
%

Fig. \ref{fig_eos_1} shows the Equation of State (EOS, or phase diagram) of the gas. The fraction of gas in the rarefied, diffuse and dense phases is $44\%$, $34\%$ and $22\%$; these phases contain $5\%$, $25\%$ and $70\%$ of the metals, respectively. Thus, while the gas mass is preferentially located in the lower density phases, metals are mostly found in dense gas, i.e. star forming regions/MC. Additionally only $\sim 30\%$ of the considered volume shows $Z>10^{-3}\zsun=Z_{\rm floor}$, i.e. it has been polluted by stars in the simulation. We note that the EOS in the $n$-$T$ plane is fairly consistent with the one found in other high-$z$ galaxy simulations \citep[e.g. see Fig. 5 in][]{wise:2012radpres}. Comparison between the EOS in the $n$-$T$ and $n$-$P$ plane highlights the relative importance of different feedback types.

The \emph{rarefied} gas is characterized by long cooling times. Thus, once engulfed by shocks, such phase becomes mildly enriched ($\langle Z\rangle \sim10^{-2}\zsun$) and remains hot ($T\sim10^{6}{\rm K}$). The enriched rarefied gas preferentially populates the $n\simeq 10^{-3}\cc$ and $T\simeq10^{6.5}{\rm K}$ region of the phase diagram. However, part of the rarefied gas has $T\simeq10^{4}{\rm K}$. This gas component has a temperature set by the equilibrium between adiabatic cooling and the photo-heating by the UV background; it feeds the accretion onto Dahlia, but it is not affected by stellar feedback. As such it is not central in the present analysis.

The \emph{dense} gas is mostly unaffected by shocks and it is concentrated in the disk. Typically, such gas has $n\sim 10^2 \cc$ and $T\sim10^{2}{\rm K}$, thus a thermal pressure $P_{\rm th}/k \sim 10^4 \cc\,{\rm K}$ is expected. However, the total gas pressure is $P/k \sim 10^7 \,\cc$ K (see the $P$-$n$ EOS). The extra contribution is provided in kinetic form by radiation pressure, thanks to the strong coupling with the gas allowed by the high optical depth of this phase. This leads to the important implication that the central structure of {Dahlia} is radiation-supported (see also Sec. \ref{sec_final_results}).

The \emph{diffuse} gas acts as an interface between the dense disk gas and the rarefied gas envelope. Diffuse gas is found both in hot ($T\sim10^{5}{\rm K}$) and cold ($T\sim10^{3}{\rm K}$) states. The cold part has a sufficiently high mean metallicity, $Z\sim 0.1\zsun$, to allow an efficient cooling of the gas. This is highlighted by the metal-weighted EOS, where we can see that most of the metals present in the diffuse phase are cold.

Note that the phase diagram also shows evidence for the classical 2-phase medium shape for pressures around $P/k \sim 10^3 \cc\,{\rm K}$, while at higher (and lower) pressures only one stable phase is allowed; nevertheless, at any given pressure a range of densities can be supported. Such situation, though, is highly dynamic and does not correspond to a true thermal equilibrium.

A final remark is that by $z=6$ a $n-Z$ correlation is already in place, although considerable scatter is present. The relation gets steeper at large densities, and at the same time the scatter decreases. Such relation arises from the superposition of the analogous relation for metal bubbles of individual galaxies ({Dahlia} and satellites). The scatter instead results from the fact that the slope of the $n-Z$ relation depends on the $SFR$ history (for an in-depth analysis see \citetalias{pallottini:2014_sim}). The average $n-Z$ relation found is consistent with the results from $z\simeq3$ galaxies \citep{shen:2014}.

\subsection{Additional ISM properties}\label{sec_final_results}

\begin{figure*}
\centering
\includegraphics[width=0.32\textwidth]{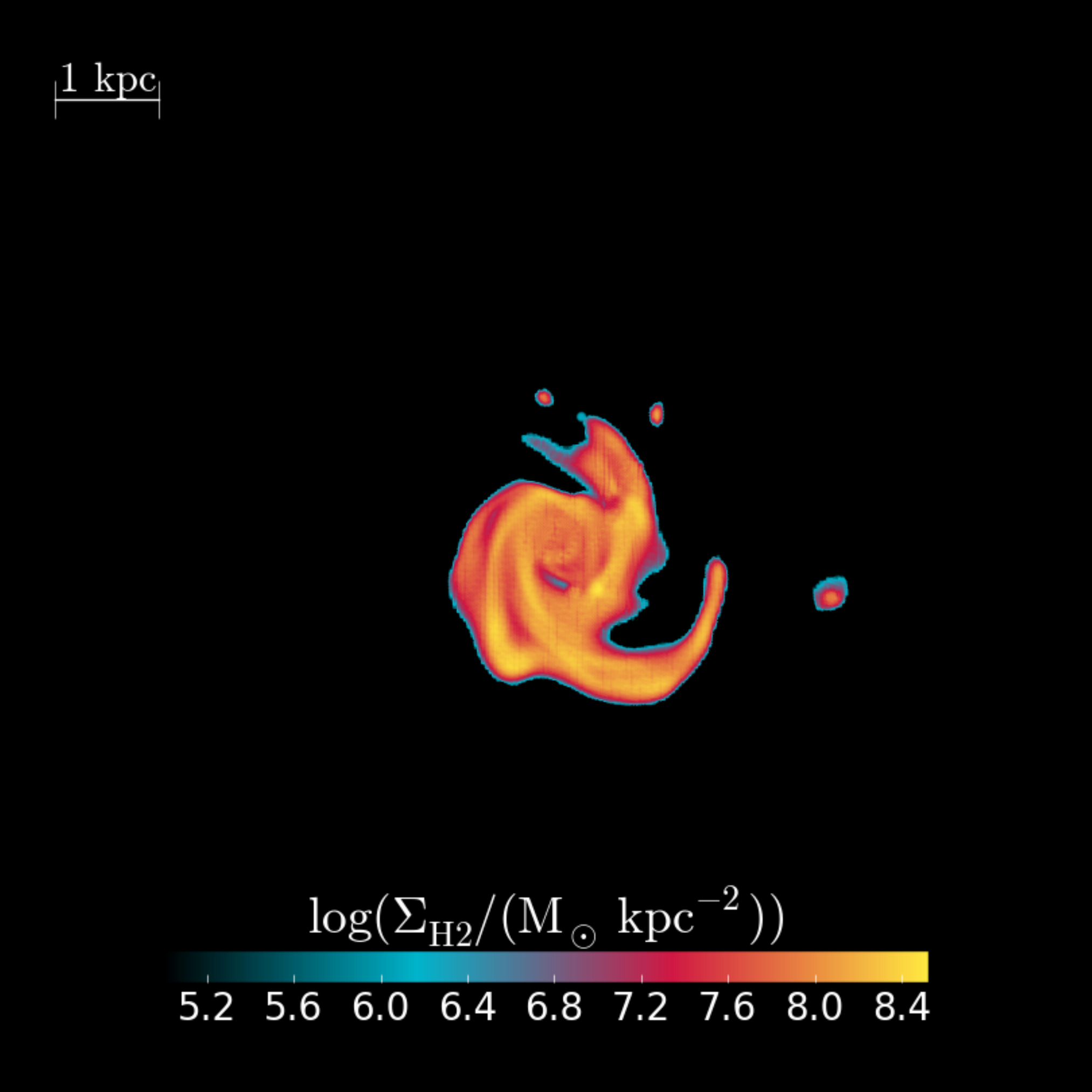}
\includegraphics[width=0.32\textwidth]{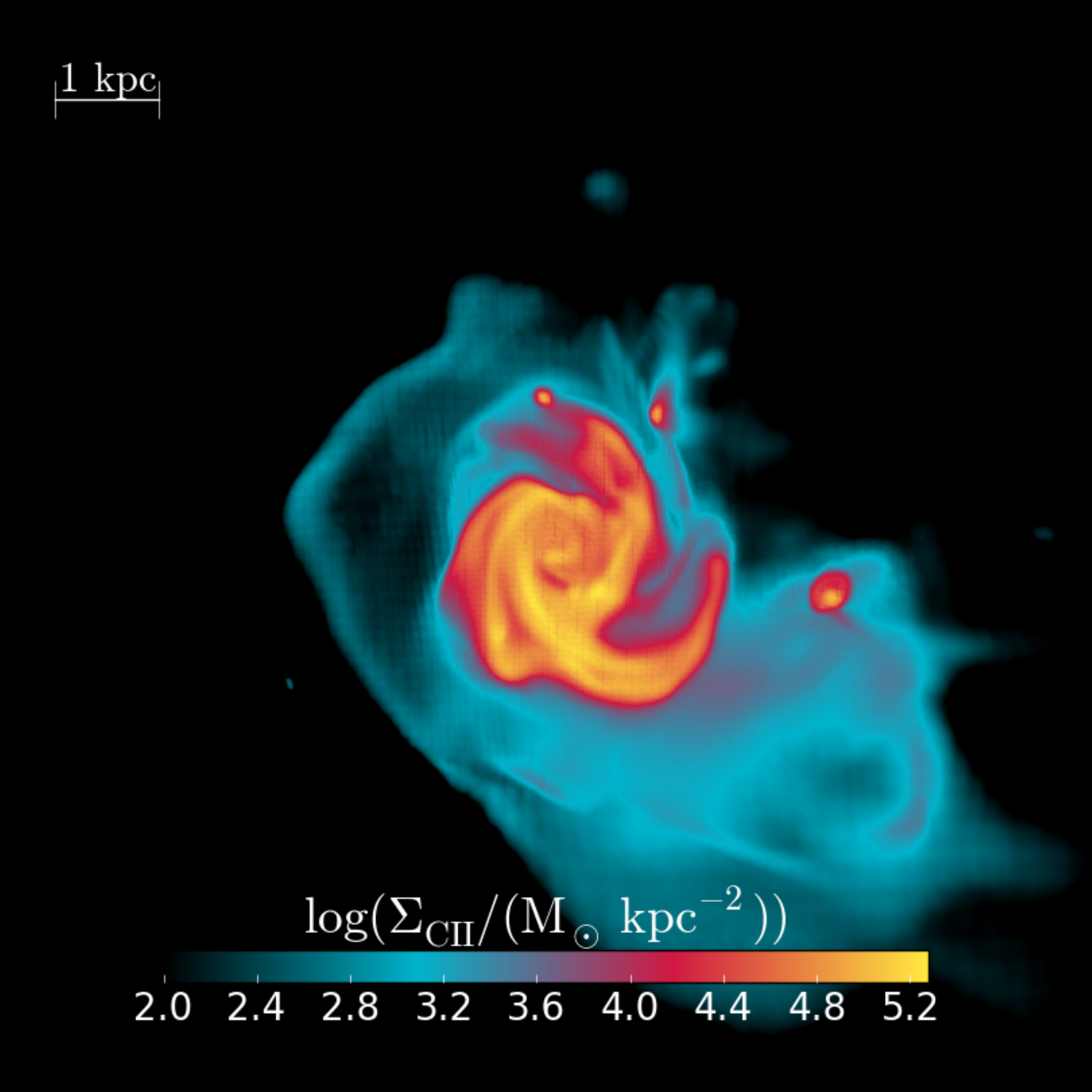}
\includegraphics[width=0.32\textwidth]{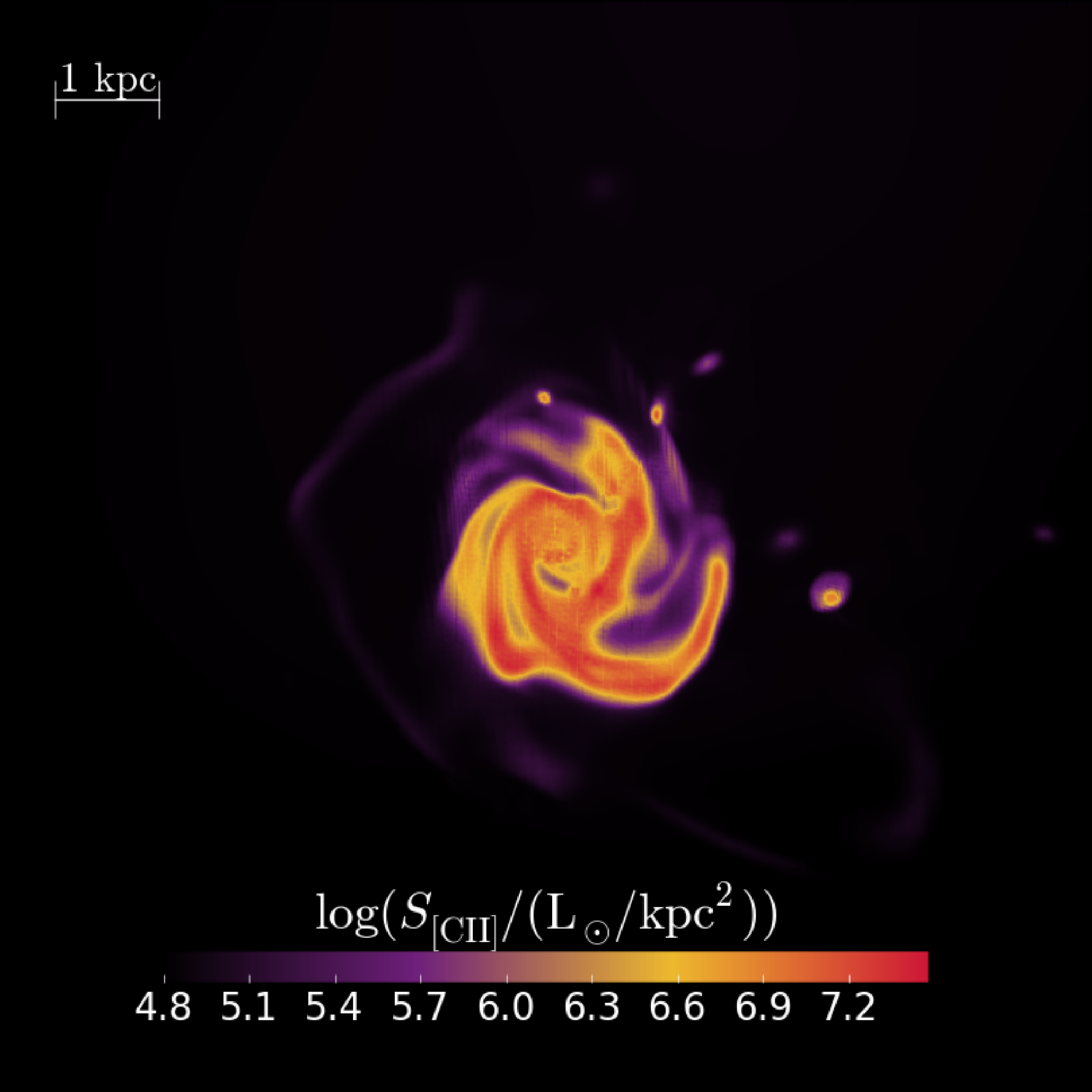}
\vspace{.5pt}

\includegraphics[width=0.32\textwidth]{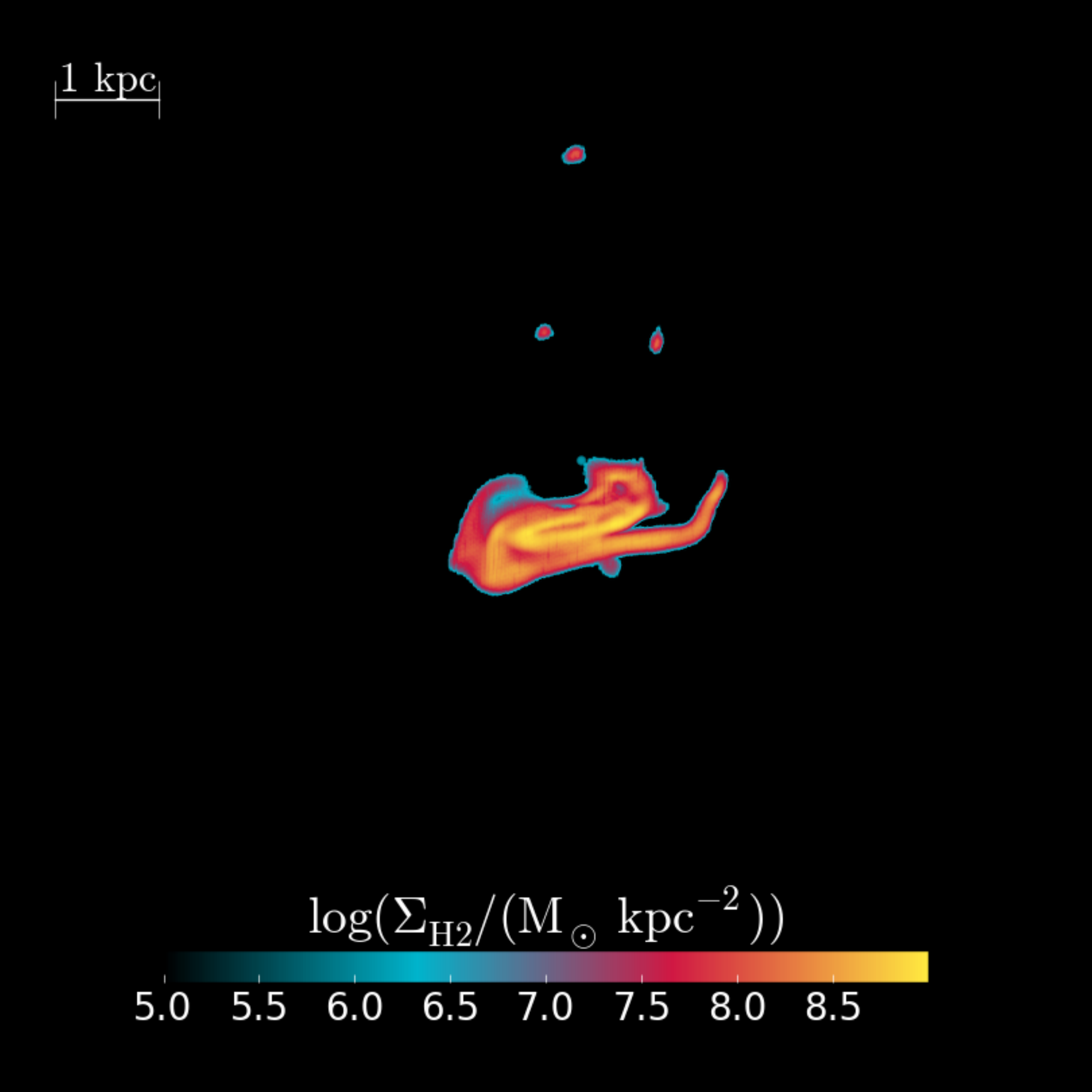}
\includegraphics[width=0.32\textwidth]{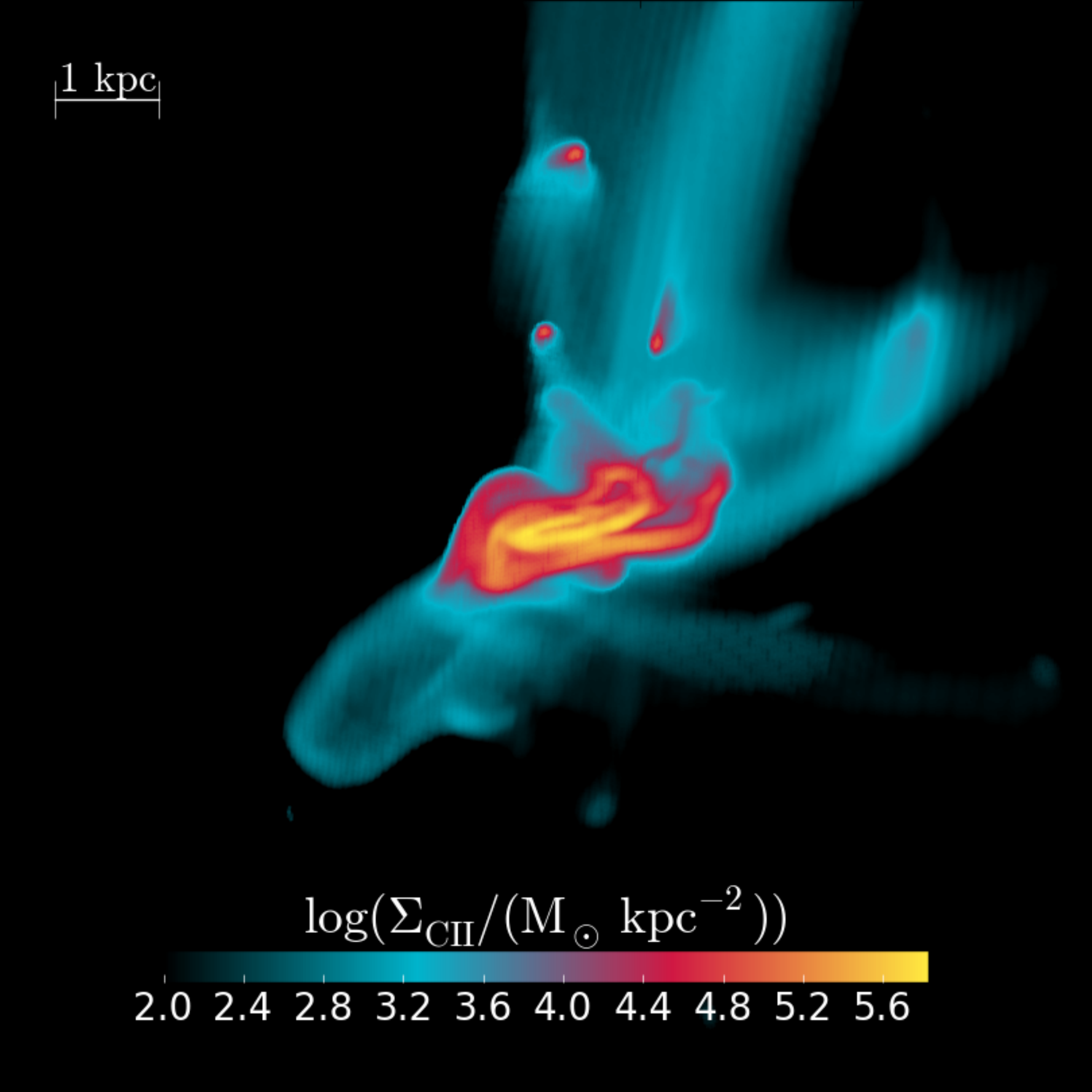}
\includegraphics[width=0.32\textwidth]{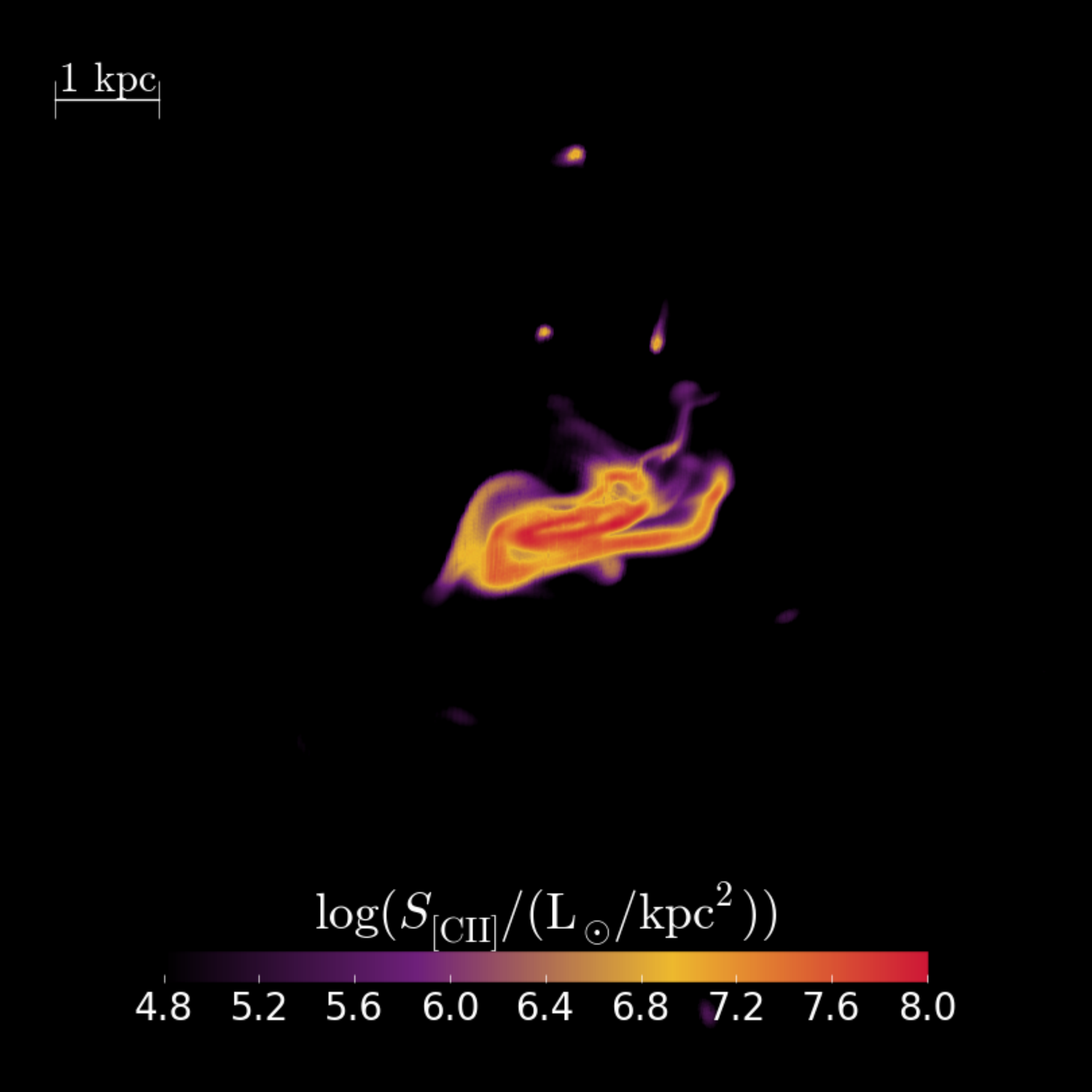}
\caption{
Face-on ({\bf upper panels}) and edge-on ({\bf lower panels}) $z=6$ {Dahlia} surface maps for \HH~density ($\Sigma_{\rm H2}/(\surfd)$ {\bf left panels}), \CIIion~density ($\Sigma_{\rm CII}/(\surfd)$ {\bf middle panels}), and \CII~brightness ($S_{\rm [CII]}/(\surfl)$ {\bf left panels}). The scale is $10$~kpc, as in the right-most panels of Fig. \ref{fig_mappe_hydro} (Sec. \ref{sec_small_scale}). Note that lower limits for the maps are drawn for visualization purposes ($\log(\Sigma_{\rm H2}/(\surfd))\simeq \log(S_{\rm [CII]}/(\surfl)) \simeq 5$, $\log(\Sigma_{\rm CII}/(\surfd))\simeq 2$). Additionally, an average of the maps is plotted in Fig. \ref{fig_mappe_results_profili}.
\label{fig_mappe_tutte}
}
\end{figure*}

\begin{figure}
\centering
\includegraphics[width=0.49\textwidth]{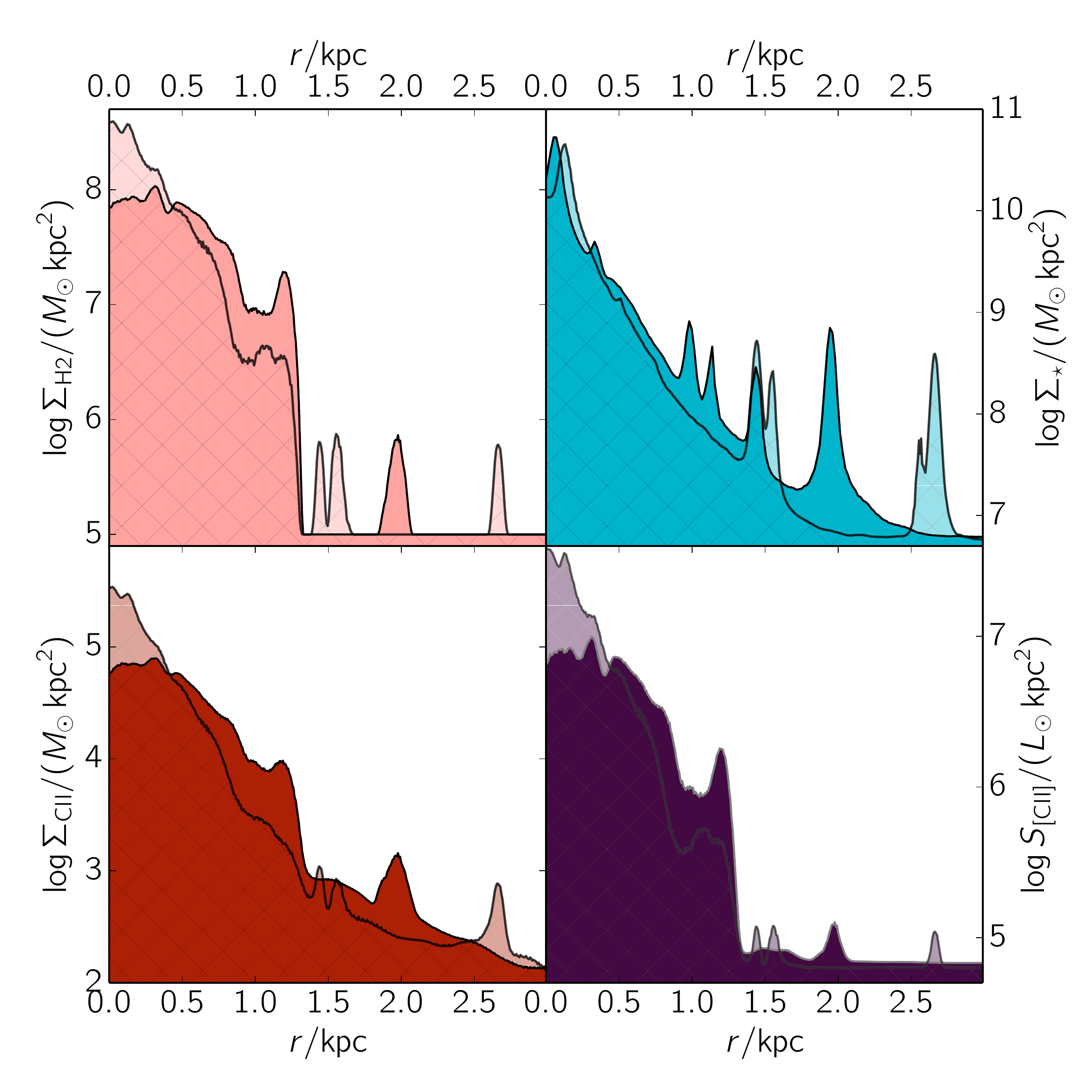}
\caption{
Radially-averaged profiles for face-on (full colour) and edge-on (transparent and hatched) views of Dahlia at $z=6$. {\bf Upper left:} 
\HH~surface density; {\bf upper right:} stellar surface density; {\bf lower left:} \CIIion~surface density; {\bf lower right:} \CII~surface brightness.
\label{fig_mappe_results_profili}
}
\end{figure}

\begin{table}
\centering
\begin{tabular}{lccc}
\hline
~ & \multicolumn{2}{c}{$r_{1/2}/{\rm kpc}$} & approximate\\
~ & face-on & edge-on & value at $r_{1/2}$\\
\hline				
\HH	& 0.59 & 0.36 &	$10^{8.23}\msun$\\
\CIIion	& 0.64 & 0.38 &	$10^{5.14}\msun$\\
stars	& 0.37 & 0.23 &	$10^{9.89}\msun$\\
\CII	& 0.60 & 0.36 &	$10^{7.25}\lsun$\\
\end{tabular}
\caption{
Summary of the effective radius ($r_{1/2}$) for the \HH, \CII, \CIIion~and stellar component in {Dahlia} at $z=6$ for the face-on and edge-on case and corresponding values for the mass/luminosity. For each entry, $r_{1/2}$ is defined as the radius including half of the mass/luminosity. The quoted radii have a reference error of $\pm 0.015\,{\rm kpc}$. Note that the approximate values of masses/luminosity at $r_{1/2}$ are insensitive to the orientation of the projection (face-on/edge-on). The full profiles are shown in Fig. \ref{fig_mappe_results_profili}.
\label{tagella_halflightradius}}
\end{table}

We conclude our analysis by inspecting the distribution of two key ISM species, molecular hydrogen and \CIIion, along with the expected surface brightness of the corresponding $158\mu$m \CII~line. The surface maps of these quantities in Dahlia ($z=6$) are shown in Fig. \ref{fig_mappe_tutte} for the face-on and edge-on view cases. For reference, in Fig. \ref{fig_mappe_results_profili} we additionally plot the radially-averaged profiles of the same quantities, and in Tab. \ref{tagella_halflightradius} we give their typical radial scales.

\subsubsection{Molecular Hydrogen}

{Dahlia} has a total \HH~mass $M_{\rm H2}\simeq 3.6\times 10^8\msun$, that is mainly concentrated in a disk-like structure of radius $\simeq 0.6\,{\rm kpc}$ and scale height $\simeq 200\,{\rm pc}$, with a sharp cut off beyond these scales\footnote{Such scales are calculated by using the principal component analysis of the \HH~distribution around the galaxy.}. The disk has mean surface density $\langle\Sigma_{\rm H2}\rangle \simeq 10^{7.5}\surfd$, that is approximately constant with radius and presents perturbed spiral arms along which the density is enhanced by a factor $\simeq 3$. The spiral arms are less pronounced than in a more massive, MW-like galaxy (see \citetalias{semenov:2015} and \citealt{ceverino:2015}). This trend with mass has already been pointed out by \citet{ceverino:2010MNRAS}.

The disk is composed by dense ($n\gsim 25\,\cc$), enriched ($Z\simeq 0.5\,\zsun$), radiation-pressure supported gas, as already discussed. It is fed by frequent mergers driving fresh gas to the centre, and supports a star formation rate per unit area of $\simeq 15\,\msun\,{\rm yr}^{-1}\,{\rm kpc}^{-2}$, i.e. more than 1000 times the Milky Way value. Fragmentation of the disk is relatively weak \citep[cfr.][]{mayer:2016}, as indicated by smooth surface density map, and also paralleling the flat metallicity profile in the inner $\simeq\,{\rm kpc}$.
For the fragmentation of the \HH~component, we caution that this result has been obtained assuming a uniform UV interstellar field; stronger fragmentation in the \HH~distribution may occur when accounting for local radiation sources: Lyman-Werner photons from these sources might in fact locally dissociate the \HH~by generating pockets of \HI~in the distribution.

While most of the \HH~gas resides in the disk, we can clearly distinguish 3 clumps of molecular gas both in the face-on and edge-on maps. These clumps are located few kpc away from the centre, and are characterized by sizes of $\sim 150\,{\rm pc}$ and $M_{\rm H2} \sim 5\times10^6 \msun$. Such clumps are Jeans-unstable and form stars as they infall and stream through the CGM, as it can be appreciated by comparing \HH~and stellar mass profiles (Fig. \ref{fig_mappe_results_profili}). 
The stellar mass profiles also highlight the presence of 3 stellar clumps at $r\sim 1\,{\rm kpc}$ with no associated \HH. These \quotes{older} clumps share the same nature of the previous ones, but the \HH~has been already consumed and/or dispersed by the star formation activity that produced the stars present at $z=6$.

\subsubsection{Singly ionized Carbon}

The \CIIion~abundance is calculated by post-processing the simulation outputs with the photoionization code \textlcsc{cloudy} (\citealt{cloudy:2013}, and see App. \ref{sez_cloudy_model}). The result is shown in Fig. \ref{fig_mappe_tutte}. {Dahlia} contains a \CIIion~mass of $M_{\rm CII} = 2.2\times 10^5 \msun$, accounting for $\sim 50\%$ of the total metals produced. About 74\% of the \CIIion~mass is located in the dense phase, $22\%$ in the diffuse phase, $4\%$ in the rarefied phase. Note that the \CIIion~mass phase distribution differs only for $\lsim 10\%$ from the $Z$ distribution (see Tab. \ref{tagella_eos_riassunto}). The difference arises because shock-heated gas can be collisionally excited to higher ionization states. Thus, to first order, we expect the \CIIion~spatial distribution to follow the metallicity one.

The face-on \CIIion~surface density has a central maximum ($\Sigma_{\rm CII} \sim 10^5\surfd$), it gradually decreases to up to $\simeq 1.2\,{\rm kpc}$, and drastically drops to $\Sigma_{\rm CII} \lsim 10^2\surfd$ beyond that radius (see also Fig. \ref{fig_mappe_results_profili}). Thus, most of the \CIIion~is located into the disk, but a more extended envelope containing a sizable fraction of mass exists. On top of this smooth distribution, there are \CIIion~enhancements corresponding to the \HH~clumps described above. 

The \CIIion~profile is similar for edge-on and face-on case. However, the edge-on has a higher \CIIion~central density and a steeper slope. While the higher central value is obviously due to the larger column density encountered along the disk, the sharp drop is related to metal transport. As most of the star formation activity is located in the disk, metals above it can be only brought by outflows which become progressively weaker with distance. Metal outflows originating from the centre are preferentially aligned with the rotation axis, and the pollution region starting from the edge is stretched by the disk rotation and by tidal interaction with satellites.

\subsubsection{Emission from singly ionized carbon}

We finally compute the expected \CII~line emission using the same prescriptions of \citet{Vallini:2013MNRAS,vallini:2015}, as detailed in App. \ref{sez_cloudy_model}. Note that for the present work we assume uniform UV interstellar radiation. This approximation is valid in the MW, where variations around the mean field value are limited to a factor of 3. The results are plotted in Fig. \ref{fig_mappe_tutte}.

Within $1\,{\rm kpc}$ from the centre the \CII~emission structure closely follows the \CIIion~distribution, and we find $S_{\rm [CII]}/\lsun \simeq 200\, \Sigma_{\rm CII}/\msun$. At larger radii the \CII~surface brightness suddenly drops, although the peaks associated with \HH~clumps are preserved. This result holds both for the face-on and edge-on cases. 

Such behavior can be understood as follows. Take a typical MC with $n = 10^2\cc$, $Z=\zsun$, and total mass $M$. Its \CII~
luminosity is $L_{[\rm CII]}/\lsun \simeq 0.1 (M/\msun)$ \citep[]{vallini:2016a,goicoechea:2015apj}. Also, the \CII~emission is $\propto Z\,n$ for $n\lsim 10^3$, i.e. the critical density for \CIIion~collisional excitation by H atoms \citep{Vallini:2013MNRAS}. Then, 
\be\label{eq_stima_luminosita}
L_{[\rm CII]} \simeq 0.1\, \left({n\over 100\, \cc}\right) \left({Z\over \zsun}\right) \left({M\over \msun}\right)\,\lsun\,.
\ee
In the central kpc, where $n \simeq 10^2\cc$ and $Z\simeq \zsun$, the luminosity depends only on the molecular mass contained in the disk, and the same holds even for \HH~clumps outside the disk. The envelope is instead more diffuse ($n\lsim 10\cc$) and only mildly enriched ($Z\lsim10^{-1}\zsun$). As a result, its \CII~luminosity per unit mass is lower. 

The emission from this diffuse component is further suppressed by the CMB \citep[][]{dacunha:2013apj,pallottini:2015_cmb,vallini:2015}. Namely, for gas with $n\lsim 0.1\,\cc$, the upper levels of the \CII~transition cannot be efficiently populated through collisions, thus the spin temperature of the transition approaches the CMB one, and to a first order the gas cannot be observed in emission.

In summary, $\simeq 95\%$ of {Dahlia} \CII~emission comes from dense gas located in the \HH~disk. Indeed, the \CII~half light radius coincides with the \HH~half mass radius, i.e. $0.59\,{\rm kpc}$ ($0.36\,{\rm kpc}$) in the face-on (edge-on) case (see also Tab. \ref{tagella_halflightradius}). Within such radius, the molecular gas has a mass $M_{\rm H_{2}}\simeq 1.69\times10^8\msun$ and the luminosity is $L_{\rm CII}\simeq 1.78\times10^7\msun$, i.e. with a \CII-\HH~scaling ratio consistent within $15\%$ from the simple estimate in eq. \ref{eq_stima_luminosita}.

Dahlia has a total \CII~luminosity $L_{\rm CII} \simeq 3.5\times 10^{7}\lsun$; this is fainter than expected on the basis of the local \CII-$SFR$ relation \citep[$L_{\rm CII}\sim 10^8 - 10^9 \lsun$, i.e.][]{delooze:2014aa}. However, at high-$z$, such relation seems to hold only for a small subset of the observed galaxies \citet[i.e.][]{capak:2015arxiv,Willott:2015arXiv15}. The majority of the observed galaxies show a strong \CII-$SFR$~deficit, when considering both detections \citep[e.g. BDF3299, A383-5.1][]{maiolino:2015arxiv,knudsen:2016arxiv} and upper limits \citep[e.g. Himiko, IOK1, MS0451-H][]{ouchi2013,ota:2014apj,knudsen:2016arxiv}.

For {Dahlia}, the \CII-$SFR$~deficit depends on multiple factors. The main contribution from \CII~emission is in the \HH~disk, that on average has $\langle Z\rangle\simeq 0.5 \zsun$, i.e. slightly lower then solar. Additionally, the gas in the disk is efficiently converted in stars ($SFR\simeq100\msun/{\rm yr}$) and has $\langle n\rangle\simeq 25\,\cc$, thus the \CII~emission is hindered (eq. \ref{eq_stima_luminosita}). Finally, there is a marginal contribution to \CII~from the diffuse and rarefied phase: $\simeq30\%$ of \CIIion~is locked in a the low density and metallicity gas that gives a negligible contribution to \CII~emission, particularly because of CMB suppression.
%

\section{Summary and discussion}\label{sec_conclusioni}

With the aim of characterizing the internal properties of high-$z$ galaxies, we have performed an AMR zoom-in simulation of \quotes{Dahlia}, a $z\simeq6$ galaxy with a stellar mass of $M_{\star}=1.6\times10^{10}\msun$, therefore representative of LBGs at that epoch. We follow the zoom-in region with a gas mass resolution of $10^{4}\msun$ and a spatial resolution of $30\,{\rm pc}$.

The simulation contains a rich set of physical processes. We use a star formation prescription based on a \HH~dependent Schmidt-Kennicutt relation. The \HH~abundance is computed from the \citetalias{krumholz:2009apj} model (Fig. \ref{fig_kmt_test}). Using stellar evolutionary models \citep{padova:1994,starburst99:1999}, we include chemical, radiative and mechanical energy inputs, accounting for their time evolution and metallicity dependence on the stellar population properties (Fig. \ref{fig_gamete_tables}). We include feedback from SN, winds and radiation pressure with a novel, physically motivated coupling scheme between gas and stars. We also compute \CIIion~abundance and the $158\mu$m \CII~emission, by post-processing the outputs with \textlcsc{cloudy} \citep{cloudy:2013}, and a FIR~emission model drawn from radiative transfer numerical simulations \citep{Vallini:2013MNRAS,vallini:2015}.

The main results can be summarized as follows:

\begin{itemize}

\item[\bf 1.] {Dahlia} sits at the centre of a cosmic web knot, and accretes mass from the intergalactic medium mainly via 3 filaments of length $\simeq 100\,{\rm kpc}$ (Fig. \ref{fig_mappe_hydro}). Dahlia has $\sim 6$ major satellites ($M_{\star}\lsim 10^{9}\msun$) and is surrounded by $\sim 10$ minor ones ($M_{\star}\sim 10^{5}\msun$). The latter represent molecular cloud (MC) complexes caught in the act of condensing as the gas streams through the circumgalactic medium (Fig. \ref{fig_sph_profile}). {Dahlia} dominates both the stellar mass ($M_{\star}\sim 10^{10}\msun$) and the SFR of the galaxy ensemble ($SFR\simeq 100\,\msun\,{\rm yr}^{-1}$, Fig. \ref{fig_sfr_smf_energy}).

\item[\bf 2.] Only a small fraction of the available energy produced by stars couples to the gas, as energy is mostly dissipated within MCs where the stars reside. Radiation dominates the feedback energy budget by a factor $> 100$ (Fig. \ref{fig_feedback_vs_time}). 

\item[\bf 3.] By $z=6$ {Dahlia} forms a \HH~disk of mass of $M_{\rm H2}= 3.6\times 10^{8}\msun$, effective radius $0.6\,{\rm kpc}$, and scale height $200\,{\rm pc}$ (Fig. \ref{fig_mappe_tutte}). The disk is dense ($n\gsim 25\,\cc$), enriched ($Z\simeq 0.5\,\zsun$), and it is fed by frequent mergers driving fresh gas to the centre, and supports a star formation rate per unit area of $\simeq 15\,\msun\,{\rm yr}^{-1}\,{\rm kpc}^{-2}$. 

\item[\bf 4.] The disk is mostly unaffected by SN shocks, and it is pressure-supported by radiation. SN/winds drive hot metal outflows (Fig. \ref{fig_eos_1}), that are either preferentially aligned with the galaxy rotation axis, or start at the edge of the disk.

\item[\bf 5.] The total \CII~luminosity of {Dahlia} is $10^{7.55}\lsun$, and $\simeq 95\%$ of the emission is co-located with the \HH~disk (Fig. \ref{fig_mappe_results_profili}). The diffuse, enriched material surrounding {Dahlia} contains $30\%$ of the \CIIion~mass, but it negligibly contributes to the \CII~emission (Fig. \ref{fig_mappe_tutte}) due to its low density ($n\simeq 10\,\cc$) and metallicity ($Z\simeq10^{-1}\zsun$). {Dahlia} is under-luminous with respect to the local \CII-$SFR$ relation; however, its luminosity is consistent with upper limits derived for most $z\sim6$ galaxies. 
\end{itemize}

We find clear indications that the SF subgrid prescription might considerably affect the \CII-$SFR$ relation and the ISM structure, as noted also by \citep{hopkins:2013arxiv}. This is because stars form in gas of different densities depending on the chosen prescription. 
In our simulation gas is converted into stars with an efficiency $\zeta_{\rm sf}\,f_{\rm H2}$, where the \HH~fraction is computed from the \citetalias{krumholz:2009apj} model and we set $\zeta_{\rm sf}=0.1$. In \citetalias{semenov:2015} the SF follows a \textit{total} (i.e. not molecular) density Schmidt-Kennicutt relation. Further the SF efficiency depends on the free-fall time and the turbulent eddy turnover time. The SF relation is derived from an empirical fit to MC simulations \citep{padoan:2012}, with no notion of the local metallicity. 

Interestingly, although the approaches are considerably different, the resulting efficiencies are compatible: in \citetalias{semenov:2015} the bulk of the star forming gas has $n\sim 10^{1.5}\cc$, as in {Dahlia} (Fig. \ref{fig_cfr_semenov}). However, with respect to \citetalias{semenov:2015}, Dahlia misses part of the very dense, star forming gas, and its corresponding contribution to \CII~from $Z\sim\zsun$ MCs with $n\sim10^{3}\cc$. These MC are expected to have high \CII~fluxes (see eq. \ref{eq_stima_luminosita}), but their abundance might be low \citep{padoan:2012}.
Further investigation is needed before we draw any solid conclusion. To this aim, we plan to upgrade our simulations to a more sophisticated non-equilibrium \HH~evolution model. This is because the chemical equilibrium assumed in \citetalias{krumholz:2009apj} does not hold in low-metallicity regimes. 

Another important caveat is that we have assumed a uniform UV background. Instead, discrete sources (stellar clusters) might have a strong impact on star formation. For example, Lyman-Werner photons might locally dissociate the \HH~by generating pockets of \HI~in the gas distribution. Thus, unshielded (low dust column density) gas in the disk would contribute only marginally to the SFR.

Furthermore, a uniform UVB assumption likely leads to inaccurate computation of the ISM thermodynamic state. We find that $Z\simeq 10^{-3}\zsun$ gas with $n\gsim 10^{2}\,\cc$ has $T\simeq 10^{4}$ (Fig. \ref{fig_eos_1}), with the temperature been set by the UVB heating. However, such gas should be likely able to self-shield from the impinging UVB, whereas internal radiation sources could still play a role \citep[e.g.][]{gnedin:2010}. 

Finally, local FUV flux variations can change the \CII~emission from individual regions of the galaxy. Also, very high FUV fluxes can photoevaporate MC on short time scales ($\lsim t_{\rm ff}$ for gas with $Z\sim 10^{-2}\zsun$, \citealt[][]{vallini:2016a}). This effect are particularly important, as it might be responsible for the displacement between the \CII~and the UV emitting region observed in BDF3299 \citep{maiolino:2015arxiv}, and in some of the \citet{capak:2015arxiv} galaxies. To solve these problems, a multi-frequency radiative transfer computation must be coupled to the present simulations. This work is ongoing and will be presented elsewhere. 


\section*{Acknowledgments}
We are grateful to the participants of \emph{The Cold Universe} program held in 2016 at the KITP, UCSB, for discussions during the workshop. 
We acknowledge the {\tt AGORA} project members and the {\tt DAVID} group for stimulating discussion. 
We thank the authors and the community of \textlcsc{pymses} for their work. 
We thank B. Smith for support in implementing \textlcsc{grackle}. 
This research was supported in part by the National Science Foundation under Grant No. NSF PHY11-25915.
S.S. was supported by the European Research Council through a Marie-Skodolowska-Curie Fellowship, project PRIMORDIAL-700907.

\bibliographystyle{mnras}
\bibliography{master}
\bsp

\appendix
\section{Blastwave model}\label{app_blastwave}

In this Sec. we present the model used to calculate the kinetic ($f_{\rm kn}$) and thermal ($f_{\rm th}$) energy fractions that a gas cell acquires during a SN explosion/because of a stellar wind (see eqs. \ref{sn_energy_gas_equation}).

First, we consider a SN explosion. Then, the picture is the following \citep[e.g.][]{cioffi:1988apj,ostriker:1988rvmp,walch:2015mnras}. In the first stage of the SN blast the stellar ejecta follow a free expansion solution, that ends when the blast have swept a mass of material roughly the mass of the ejecta, $\sim 1 \msun$ per SN. Then, the shock enters in a Sedov-Taylor stage (ST). The ST stage is adiabatic and the available energy is divided in thermal and kinetic parts, that accounts for $f_{\rm th}\simeq 0.7$ and $f_{\rm kn}\simeq 0.3$ of the total\footnote{Such values of $f_{\rm th}$ and $f_{\rm kn}$ are calculated by assuming $\gamma=5/3$ for the gas adiabatic index. See Tab. III in \citetalias{ostriker:1988rvmp} for the general solution.}, respectively. The ST stage ends when radiative losses becomes important, at $t\simeq t_{\rm cool}$, the cooling time of the ambient gas. At the contact discontinuity the gas cooling causes the formation of a thin shell (shell formation stage, SF), and after that the shock proceeds snowplowing through the ambient medium, driven by the pressure of the gas interior (pressure driven snowplow, PDS). When all the thermal energy is radiated away, then the blastwave enters in the so called momentum conserving snowplow (MCS). In the MCS stage the momentum is conserved, while the remaining energy, purely kinetic, is gradually lost because of the work done by the blast on the ambient material, and eventually the blast stops.

In each stage, the blastwave can be modelled by following the analysis by \citet[][hereafter \citetalias{ostriker:1988rvmp}]{ostriker:1988rvmp}. A self similar blastwave can be described by the evolution of the shock front at time $t$ as $r_{s}\propto t^{\eta}\,,$ where $\eta$ is a constant that determines the type of blast (ST, MCS, $\dots$). The velocity of propagation of the shock is $v_{s} = \eta r_s / t$, and, using the virial theorem (see eq. 3.3 in \citetalias{ostriker:1988rvmp}), the total energy of a blastwave can be written as
\be
\label{eq_energy_shock}
E(t) = (4\pi/3) \sigma_{\eta} \rho\, r_{s}^3 v_s^{2}\propto t^{5\eta-2}\,,
\ee
where $\sigma_{\eta}$ is a dimensionless constant and $\rho$ is the density of the material swept by the blast. Within the presented formalism (eq. \ref{sn_energy_gas_equation}), $f=E(t)/E_{0}$, where $E_{0}$ is the input energy, i.e. $E_{0}=[\epsilon_{\rm sn}(t_{\star}+\Delta t)-\epsilon_{\rm sn}(t_{\star})] M_{\star}$ (eq. \ref{eqs_def_mec_energy}).

The cooling time is critical in calculating the evolution of the blast between different stages. Here the cooling time $t_{\rm cool}$ is defined as the time $t$ when $t = K_{\rm b} T_{s}/(\rho_{s}/m_{\rm p} \Lambda)$, where $K_{\rm b}$ is the Boltzmann constant, $T_{s}=T_{s}(t)$ and $\rho_{s}$ are the temperature and density at the shock front, $m_{\rm p}$ the proton mass and $\Lambda=\Lambda(T_{s},Z)$ the cooling function.

Note that calculating the cooling function with \textlcsc{grackle} or analytical approximations \citep[e.g.][]{raymond:1976apj,nisikawa:1997astro,koyama:2002} yield comparable results. This happens because the gas start to cool at $T_{s}\gsim 10^7 K$, when Bremsstrahlung is the dominant cooling process, and it is independent of the metallicity, i.e. $\Lambda\propto T^{-1/2}$. Thus, as in \citet[][see eq 3.10]{cioffi:1988apj}, the cooling time can be approximated as $t_{\rm cool} = 3.61\,10^{-2} (E_{0}/{\rm foe})^{3/14} (\rho/m_{\rm p}\,\cc)^{4/7}\myr$ \citep[see also][in particular see eqs. 6 and 7]{kim:2015apj}.

The typical range of input energy ($1\leq E_{0}/{\rm foe}\lsim10^{3}$) and ISM density ($10^{-1}\lsim \rho/m_{\rm p}\,\cc\lsim10^{3}$), thus the cooling time is in the range $10^{-3} \lsim t_{\rm cool}/\myr\lsim 1$. Since the expected simulation time step is $\Delta t \sim 10^{-2} \myr$, we can further simplify the blastwave picture\footnote{For the complete picture of blastwave evolution, we refer the reader to \citetalias{ostriker:1988rvmp}, in particular Fig.1, table IV and references therein.}.

The free expansion stage is shorter than our typical simulation time step ($ t\lsim 10^{-4} \myr$, e.g. see eq. 1 in \citealt{kim:2015apj}), thus we assume that the SN starts in the ST stage ($\eta=2/5$). After $\simeq t_{\rm cool}$ the energy of the shock is roughly half the initial and the blastwave is in the PDS stage ($\eta=2/7$). In both stages, the total energy is given by the sum of kinetic and thermal terms, and the relative fraction of kinetic ($f_{\rm kn}$) and thermal ($f_{\rm th}$) energies are constants (see eqs. 3.16 and 3.18 in \citetalias{ostriker:1988rvmp}) that depends on $\eta$ and on the internal structure of the blastwave. In the intermediate SF stage we approximate $f_{\rm kn}$ and $f_{\rm th}$ by linearly interpolating between the ST and initial PDS values. The time when the blast enters in the SF and PDS stages can be calculated following the analytical fit to the simulation presented in \citet{cioffi:1988apj}, i.e. $0.14\, t_{\rm cool}$ and $0.4\, t_{\rm cool}$ for SF and PDS, respectively (see their eq. 3.15). An example of such model is presented in Sec. \ref{sezione_blast} (in particular, see Fig. \ref{fig_blastwave_sketch}). Note that the model is consistent with the result from SN exploding in an homogeneous medium, however a blastwave propagating in an inomhogeneous medium behave differently, since the blast travels unimpeded through path of lower density medium \citep[][see in particular the fit in eq. 9 and 10 and the different scalings given in eq. 11 and 12]{martizzi:2015mnras}. We will explore this aspect in a future work.
%

The winds can be treated within the same blastwave formalism (see Sec. VII in \citetalias{ostriker:1988rvmp}). The stage evolution of the winds is similar to the SN one: first the wind is adiabatic, then the outer shock begin to radiate away and a thin shell is formed, and eventually the shock becomes momentum conserving. However in the wind case the energy injection is not impulsive, as the stars input a continuous energy injection with a (roughly) constant luminosity. Thus, we have a different blastwave structure and consequently $f_{\rm th}$ and $f_{\rm kn}$ in each stage.

The wind solution is given by \citet[][hereafter \citetalias{weaver:1977apj}]{weaver:1977apj}. The structure can be spatially divided in three parts \citepalias[see Fig. 1 in][]{weaver:1977apj}: near to the stars we have the stellar wind ($a$), and the gas is contained in the region of shocked stellar wind ($b$) and in the shell of shocked gas ($c$).

In the adiabatic case we have a (roughly) constant wind luminosity ($L$) and no radiative losses, thus dimensional analysis implies $\eta=3/5$ in eq. \ref{eq_energy_shock}. In this stage the gas in region $b$ has $5/11$ of the total energy, that is in a purely thermal form, while the gas in $c$ contains energy in both thermal and kinetic form. Summing the contribution of both regions, \citetalias{weaver:1977apj} finds $f_{\rm th} \simeq 0.78$ $f_{\rm kn} \simeq 0.22$ (see also eqs. 7.11 in \citetalias{ostriker:1988rvmp}). Note that the thermal energy fraction is larger to the corresponding one in the ST stage of the SN driven blast.

In the next stage radiative losses becomes important ($t>t_{\rm cool}$), and region $(c)$ collapses to a thin shell. All the energy ($5/11\,L\, t$) would be contained in the gas in $b$ in thermal. The physical situation is that the injected fluid is adiabatic, while in the ambient radiative losses would be dominant. This happens because the density of the injected fluid is much less then the ambient medium, and this consequently affects the the cooling time scales. However, the modellization of this stage cannot be readily implemented as a subgrid model in our simulation, because we we cannot easily keep track of the internal structure of the cells in the simulation.

In our code we opt to go directly from the adiabatic stage ($f_{\rm th} = 0.78$, $f_{\rm kn} = 0.22$, \citetalias{weaver:1977apj}) to the momentum conserving stage when $t>t_{\rm cool}$. The latter is described by $f_{\rm kn}=1$, and $E(t)$ is given by eq. \ref{eq_energy_shock} with $\eta = 1/4$ (see eq. 7.20 in \citetalias{ostriker:1988rvmp}).

As noted in \citet[][]{agertz:2012arxiv,walch:2015mnras,fierlinger:2016,kortgen:2016}, we expect wind injection to make the SN more efficient, since SN blast sweeps through a gas with a lower density, thus with a longer cooling time. This point applies both to wind and radiation pressure injection.

Finally, note that in the blastwave modellization, we have neglected the instabilities that can arise during the thin shell formation stage \citep{madau:2001apj,mcleod:2013mnras,badjin:2015}, and we do not explicitly consider the effect of multiple blastwave events \citep{walch:2015mnras,fierlinger:2016}. This issues will be addressed by future work.

\section{Radiation pressure: optical depth and IR-trapping}\label{app_rad_press}

The radiation pressure is dependent on the rate of momentum injection ($\dot{p}_{rad}$, eq. \ref{eq_rad_moment_injection}), that in turn depend on the optical depth to ionizing photons ($\tau_{\rm ion}$) and the IR-trapping of the UV radiation ($f_{\rm ir}$).

For each gas cell, the optical depth $\tau_{\rm ion}$ can be calculated by averaging the ionization cross section on the stellar spectra assumed in our model (see eqs. \ref{eqs_def_rad_energy}), that are taken from \textlcsc{starburst99} \citep{starburst99:1999,starburst99:2010apjs}. Namely, $\tau_{\rm ion} = \sigma_{\rm ion} N_{H}$, where ionization cross section $\sigma_{\rm ion}$ is obtained as a Rosseland mean
\begin{subequations}
\be
\sigma_{\rm ion} = \int_{912\,\angstrom}^{4000\,\angstrom} L_{\lambda}\sigma_{\rm ion}^{\lambda}\,{\rm d}\lambda / L_{\rm ion}\,,
\ee
where the cross section as a function of wavelength is given by \citet[][]{osterbrock:1989book}
\begin{align}
\sigma_{\rm ion}^{\lambda} = 6.3\times 10^{-18} (&1.34\,(912\,\angstrom/\lambda)^{-2.99} +\\
-& 0.34\,(912\,\angstrom/\lambda)^{-3.99}). \nonumber
\end{align}
\end{subequations}
For the IR-trapping we assume $f_{\rm ir} = \tau_{\rm ir}$, the infrared dust optical depth, similarly to \citet{agertz:2012arxiv}. As noted in \citet{krumholz:2012radpress}, the assumption $f_{\rm ir} = \tau_{\rm ir}$ is approximately correct for small values of the optical depth, i.e. $\tau_{\rm ir}\lsim 20$. This is our case, since in simulations with resolution scale larger than $10\,{\rm pc}$, we expect $\tau_{\rm ir}\lsim 10$ \citep{rosdahl:2015mnras}. As reported in Sec. \ref{sec_feedback_res}, $\tau_{\rm ir}\lsim 10^{-1}$ in our simulation (see also Fig. \ref{fig_feedback_vs_time}).

Using the cross section from \citet[][see also \citealt{semenov:2003aa}]{draine:2003araa}, we can write
\begin{subequations}
\be
 \tau_{\rm ir} \simeq 1.79\times10^{-24} (N_{H}/{\rm cm}^{-2}) (Z/\zsun) (T_{\rm dust}/100\,{\rm K})^{2}\,,
\ee
where $T_{\rm dust}$ is the dust temperature. $T_{\rm dust}$ can be calculated as \citep{dayal:2010mnras}
\be
 T_{\rm dust} = 6.73\, (L_{\rm ir}/\lsun)^{1/6} (M_{\rm dust}/\msun)^{-1/6}\,,
\ee
where $L_{\rm ir}$ is the reprocessed UV luminosity, i.e. $L_{\rm ir} = L_{\rm uv} (1-\exp(-\tau_{\rm uv}))$, and $M_{\rm dust}$ is the dust mass, i.e.
\be\label{eq_dust_mass}
 M_{\rm dust} = \mathcal{D}_{\odot} \rho V_{cell} (Z/\zsun)\,,
\ee
with $\mathcal{D}_{\odot} = 6\times10^{-3}$ being the solar dust to gas ratio \citep[e.g.][]{hirashita:2002mnras}.

We account for CMB heating on dust, that can be important at high redshift \citep[e.g.][]{dacunha:2013apj}. The CMB heating is calculated by using a correction to the dust temperature \citep[][in particular see eq. 12]{dacunha:2013apj}, i.e.
\be
 T^{\rm corr}_{\rm dust}(z) = (T_{\rm dust}^{4+\beta} +T_{\rm CMB}^{4+\beta}(z) -T_{\rm CMB}^{4+\beta}(z=0))^{1/(4+\beta)}\,
\ee
where $T_{\rm CMB}(z)= 2.725\,(1+z)$ is the average CMB temperature at $z$, and $\beta$ is the dust emissivity coefficient \citep[][]{draine:1984apj}. Similarly to \citet[][]{dacunha:2013apj}, we assume a fiducial dust emissivity of $\beta =2$.
\end{subequations}

Note that dust can be destroyed by sublimation or evaporation. In the adopted chemical network (and the current \textlcsc{grackle} version), dust abundance is not accounted self consistently, and we calculate it by assuming that the $M_{\rm dust}$ is proportional to the metal mass of the gas (eq. \ref{eq_dust_mass}). To mimic dust destruction, we assume a negligible dust contribution to the optical depth when $T_{\rm dust}>2\times 10^{3}$ \citep[e.g.][]{bauer:1997aa}. We neglect dust sputtering driven by SNs \citep[e.g.][]{valiante:2009mnras,draine:2011apj}, which might lead to dust destruction, and thus further reduce the radiation pressure efficiency.

Finally, we note that in our radiation feedback model (eq.s \ref{eq_rad_moment_injection} and \ref{eq_red_press_energy_increase}), energy conservation is guaranteed by construction, if only the UV and ionizing contributions were present. However, the IR-trapping is accounted by approximate formulas ($f_{\rm ir} = \tau_{\rm ir}$), that are consistently evaluated with the gas properties but do not account for radiative losses in the IR cascade. For this reason, we have added the additional energy conservation check presented in Sec. \ref{sec_rad_press}. Note that a posteriori we have found that IR-trapping is subdominant contribution to radiation pressure ($\tau_{\rm ir}\lsim10^{-1}$, Sec. \ref{sec_feedback_res}), thus probably the check is not necessary.

\section{Postprocessing ionization state and emission}\label{sez_cloudy_model}

Radiative transfer is not followed during the evolution of the simulation. To compute the ionization state of the various atomic species, we post-process the simulation outputs using the photoionization code \textlcsc{cloudy} \citep{cloudy:2013}. We consider a grid of models based on the density ($n$), temperature ($T$) and metallicity ($Z$) of the gas in our simulation. We produce a total of $10^3$ models, that are parameterized as a function of the column density ($N$).

The radiation fields includes the UVB intensity at $912\, {A}$ \citep[][]{Haardt:2012}, the CMB background and a galactic background, that is obtained by rescaling with {Dahlia} $SFR$ the Galaxy spectrum \citep{black:1987}, in particular $G$ the FUV flux in the Habing band, that is usually normalized to the Galactic mean value $G_{0}$. Thus we use a use $G=130\,G_{0}$ in our calculation. Note that larger value of $G$ does not yield a large variation of the expected \CII~in molecular gas \citep[][]{vallini:2016a}.

Note that accounting for the UVB is not relevant for the ionization state of the gas in the proximity of galaxies \citep[][]{gnedin:2010}: because of the high density environment, the gas is efficiently shielded and the galactic emission is the dominant radiation source. Thus, in our \textlcsc{cloudy} models we consider that the gas is shielded by a column density of $N\simeq 10^{20}{\rm cm}^{-2}$.

Spatial variation of the incident radiation are are not considered in the present work. It is to note that the variation of $G$ is very low in our Galaxy \citep{habing:1968,wolfire:2003apj}, i.e. $\langle G\rangle =G_0$ and $(\langle (G/G_0)^{2} - 1\rangle )^{1/2}\simeq 3$. However, in the close proximity of young OB associations, the flux can be very high, e.g. $10^6 G_{0}$ at $0.1\,{\rm pc}$ from a single OB star \citep{Hollenbach:1999}, or -- equivalently -- $10^6 G_{0}$ at $10\,{\rm pc}$ from a starburst with $SFR\simeq \msun/{\rm yr}$. As the flux propagate with $1/r^2$, such variation are not perceived in most of the volume. Such effect is not considered in the present work.

Using the \CIIion~density obtained with \textlcsc{cloudy}, we compute the \CII~luminosity by using eq. 3 in \citet{Vallini:2013MNRAS}. The effect of CMB suppression of \CII~is included in the model \citep[][]{pallottini:2015_cmb,vallini:2015}. Such effect suppress the emission where the spin temperature of the \CII~transition is close to the CMB one. This is relevant for low density ($n\lsim 10^{-1}\,\cc$) medium, that does not have enough collision to decouple from the CMB. We note that the result from the present \CII~modelling is consistent with the one obtained by directly using \textlcsc{cloudy} to compute the PDR emission.

Note that the current model does not account for the photoevaporation effect on MC, that has an important impact on FIR emission \citep[][]{vallini:2016a}, particularly when including a spatially varying FUV field. More detailed modelling will be accounted in a future work.

\label{lastpage}

\end{document}